\documentclass{aa}  
\usepackage{lscape}
\usepackage{longtable,threeparttable}
\usepackage{graphicx}
\usepackage{txfonts}
\usepackage{caption}
\usepackage{subcaption}

\usepackage{hyperref}
\hypersetup{
	unicode=false,
	pdftoolbar=true,
	pdfmenubar=true,
	pdffitwindow=false,
	pdfstartview={FitH},
	pdftitle={My title},
	pdfauthor={Author},
	pdfsubject={Subject},
	pdfcreator={Creator},
	pdfproducer={Producer},
	pdfkeywords={keyword1} {key2} {key3},
	pdfnewwindow=true,
	colorlinks=true,
	linkcolor=red,
	citecolor=blue,
	filecolor=magenta,
	urlcolor=cyan
}

\begin{document}

   \title{Hot subdwarfs in close binaries observed from space I: \\ orbital, atmospheric, and absolute parameters and the nature of their companions}

   %\subtitle{Parameters from high signal-to-noise light curves of reflection effect systems}
    \titlerunning{Close sdB binaries from TESS I}
   \author{V. Schaffenroth
          \inst{1}
          \and
          I. Pelisoli\inst{2,1}
          \and
          B.~N. Barlow\inst{3}
          \and S. Geier\inst{1}
          \and T. Kupfer\inst{4}
         % \and S. Walser\inst{3}
          }
\authorrunning{Schaffenroth et al.}
   \institute{Institute for Physics and Astronomy, University of Potsdam, Karl-Liebknecht-Str. 24/25, 14476 Potsdam, Germany\\
              \email{schaffenroth@astro.physik.uni-potsdam.de}
         \and
             Department of Physics, University of Warwick, Gibet Hill Road, Coventry CV4 7AL, UK
        \and
        Department of Physics and Astronomy, High Point University, High Point, NC 27268, USA
        \and 
        Department of Physics and Astronomy, Texas Tech University, PO Box 41051, Lubbock, TX 79409, USA
             }

   \date{Received 08 June 2022/ Accepted 03 July 2022 }

% \abstract{}{}{}{}{} 
% 5 {} token are mandatory
 
  \abstract
   {About a third of the hot subdwarfs of spectral type B (sdB), which are mostly core-helium burning objects on the extreme horizontal branch, are found in close binaries with cool, low-mass stellar, substellar, or white dwarf companions. They can show light variations due to different phenomena.}
   {Many hot subdwarfs now have space-based light curves with high signal-to-noise ratio available. We used light curves from the Transiting Exoplanet Survey Satellite and the \textit{K2}  space mission to look for more sdB binaries. Their light curves can be used to study the hot subdwarf primaries and their companions and get orbital, atmospheric, and absolute parameters for those systems, when combined with other analysis methods.}
   {By classifying the light variations and combining this with the fit of the spectral energy distribution, the distance derived by the parallaxes obtained by \textit{Gaia} and the atmospheric parameters, mainly from the literature, we could derive the nature of the primary and secondary in 122 (75\%) of the known sdB binaries and 82 newly found reflection effect systems. We derive absolute masses, radii, and luminosities for a total of 39 hot subdwarfs with cool, low-mass companions, as well 29 known and newly found sdBs with white dwarf companions. %By analysing the light curves of 19 sdB binaries showing a reflection effect we could derive for the first time for a larger sample the masses and radii of the companions. Moreover, we could constrain the masses of eight companions in sdB binaries showing a tiny ellipsoidal modulation.
   }
   {The mass distribution of hot subdwarfs with cool, low-mass stellar and substellar companions differs from those with white dwarf companions, implying they come from different populations. By comparing the period and minimum companion mass distributions, we find that the reflection effect systems all have M dwarf or brown dwarf companions, and that there seems to be several different populations of hot subdwarfs with white dwarf binaries --- one with white dwarf minimum masses around $0.4\,M_\odot$, one with longer periods and minimum companion masses up to $0.6\,M_\odot$  and at the shortest period another with white dwarf minimum masses around $0.8\,M_\odot$.
   %The analysis of several sdB binaries with cool companions allowed us to derive the masses and radii of the companions confirming them to be M dwarf companions with masses from $0.09$ to $0.4\,\rm M_\odot$ covering the whole mass range of M dwarfs. Moreover, 
   We also derive the first orbital period distribution for hot subdwarfs with cool, low-mass stellar or substellar systems selected from light variations instead of radial velocity variations. It shows a narrower period distribution from 1.5 hours to 35 hours compared to the distribution of hot subdwarfs with white dwarfs, which ranges from 1 hour to 30 days. These period distributions can be used to constrain the previous common envelope phase.% together with the masses and separations derived for those systems.
   }
   {}


   \maketitle
%
%-------------------------------------------------------------------

\section{Introduction}
Hot subdwarfs of spectral type O and B (sdO/Bs) are a mixture of different kinds of evolved stars located at or close to the bluest end of the horizontal branch, referred to as the extreme horizontal branch (EHB). Subdwarf O stars consist of many different objects including post-red giant branch and post-asymptotic giant branch stars. Most sdBs on the other hand, which are mostly found on the EHB, are core-He burning objects with very thin envelopes and masses close to the core-helium-flash mass of 0.47 $\rm M_\odot$ --- for sdBs coming from low-mass star progenitors. A higher mass range of 0.35-0.65 $\rm M_\odot$ is possible for sdBs originating from more massive stars. A small fraction of sdBs is composed of extremely low-mass  pre-white dwarfs (pre-ELM WD), which can cross the EHB on their way to the WD cooling track \citep{heber:2009,heber:2016}. Significant mass-loss on the red giant branch (RGB) is necessary to form sdO/Bs, and \citet{han:2002, han:2003} proposed different binary evolution channels to form such objects. Stable mass transfer leads to a composite sdB system with a K to F type companion and orbital periods of a few hundred days \citep{vos:2018}. They are double--lined binaries in the visible range showing spectral features from both the sdB and the cool companion. In the case of a larger mass ratio --- above 1.2-1.5 --- the mass-transfer is unstable and results in a common-envelope phase. The outcome of this poorly understood phase \citep{ivanova} is a sdB with a cool-low mass companion with a period of 0.05 day to around one day \citep{erebos}. Finally, after a stable mass transfer phase has passed, unstable mass-transfer can commence once the sdB's companion evolves into a red giant, leading to a short--period binary with a WD companion. He-core burning sdBs will evolve to sdOBs and sdOs after He-exhaustion in the core, before contracting onto the WD cooling track.

Most sdB binaries exhibit different kinds of variability in their light curves. \citet{pelisoli:2020} found that many of the composite sdB binaries show small amplitude variations in their light curves with periods of 0.5 d to a few days, due to spots on the companions. Subdwarf B stars with WD companions can show ellipsoidal deformation and even Doppler beaming in their light curves when the orbit is close enough and the WD massive enough \citep{kupfer:22, kupfer:20a, kupfer20, kupfer:2017}. 
Systems with cool, low-mass companions show unique light curve variations resulting from the extreme temperature difference and small separation distance between the two stars (as small as $0.5-1\rm\,R_\odot$). The UV--bright hot subdwarf irradiates the side of the cool companion facing it, and this leads to hot and cold sides of the companion due to their being tidally locked. The irradiated face rotating in and out of view produces a quasi--sinusoidal flux variation called the reflection effect that exhibits broad minima and sharper maxima. In systems with inclination angles $\gtrsim 60-65^\circ$, eclipses can be observed given the right combination of stellar sizes and orbital separation. Such eclipsing sdB binaries are called HW Vir systems \citep[e.g.][]{menzies:1986,erebos,Schaffenroth2020}. Finally, some hot subdwarfs show variability due to short-period pulsations on the order of minutes (for sdO/B with $T_{\rm eff}>30000\,\rm K$) and long-period (for sdO/B with $T_{\rm eff}<30000\,\rm K$), low-amplitude pulsations on the order of hours \citep[see][for a summary]{lynasgray21,kupfer2019}. Some targets in binaries can even show variability due to both pulsations and binary effects \citep[e.g.][]{nyvir}.

\citet{geier_gaia_catalog} published a catalogue of 39\,800 hot subluminous star candidates with $G<19\,\rm mag$ based on \textit{Gaia} DR2 \citep{gaia_dr2} colors, parallaxes, and proper motions and several ground-based, multi-band photometry surveys. They expect the majority of the candidates to be hot sdO or sdBs, followed by blue horizontal branch stars, hot post-AGB stars, and central stars of planetary nebulae (PN). The main purpose of their catalogue is to serve as a target list for current and future large-scale photometric and spectroscopic surveys. 

One of those surveys is the \textit{TESS} (Transiting Exoplanet Survey Satellite) mission \citep{TESS}, which is observing over 90\% of the northern and southern sky in different sectors. Each sector has a field of view of $24^\circ \times 90^\circ$ and is observed for 27 consecutive days, with a short break halfway through for data downlinking.
%which is observing different sectors with over 90\% of the northern and southern sky with a field of $24^\circ \times 90^\circ$ 27 d each with an exposure time of 2s being released every month. 
The full frame images are downloaded every 30 min (and since sector 28, every 10 min), providing light curves of all stars in the field-of-view of 30 min (10 min) cadence. A number of pre-selected stars are downloaded every 2 min (since sector 28 some additionally also with 20s cadence). As members of the  \textit{TESS} Asteroseismic Consortium (TASC) Working Group (WG) 8 on compact pulsators with the subgroup WG8.4 on binaries, we were able to provide input target lists including bright hot subdwarfs from the hot subluminous star candidate catalogue \citep{geier_gaia_catalog}, as well as with Guest Investigator programs G022141, G03221, and G04091 (PI: Brad Barlow). The majority of these targets were submitted because they either were known variable hot subdwarfs or were strong candidates for variability based off of their anomalous \textit{Gaia} flux errors and other metrics \citep{barlow22}. This provides us with a few thousand space-quality light curves of hot subdwarf stars, including the few tens of light curves already obtained from \textit{K2}  \citep{k2} from different successful proposals. Consequently, we possess for the first time an expansive, high S/N data set of hot subdwarf light curves. \citet[][]{tess_south} and \citet[][]{tess_north} used the 30 min cadence \textit{TESS} light curves of observed targets from the hot subluminous star candidate catalogue \citep{geier_gaia_catalog} to search for light variations of hot subdwarf candidates and found several sdB+dM/BD candidates.

In this paper we present our search for hot subdwarfs with cool, low mass companions showing the reflection effect and hot subdwarfs with white dwarf companions showing ellipsoidal deformation and/or beaming, as well as a characterization of these systems. In section 2 we give more details to our target selection and our search for light variations. In section 3 we present our characterization of the primary star using the parallaxes and proper motions provided by \textit{Gaia}, as well as the fit of the spectral energy distribution allowing us to get a mass distribution for the sdB in close binaries. In section 4 we show the distribution of the orbital parameters (period, semi-amplitude of the radial velocity curve) of our targets and compare the different populations. %In section 5 we present our analysis of the light curves of the sdB+WD systems and in section 6 we show the analysis of the close sdB binaries with solved orbits showing a reflection effect. 
In section 5 we conclude and provide a short summary of our results.

%We will also show a more detailed analysis of reflection effect systems with solved radial velocity (RV) curves giving masses and radii of the companions for the first time. 

\section{Target selection and search for light variations}

\begin{table}
\caption{Result of our light curve search}
\label{search}
	\begin{tabular}{ll}
		\hline\hline
		type&number (analyzed)\\
		\hline
		new reflection effect systems&82(0)\\
		reflection effect systems with solved orbits & 20 (17)\\
		HW Vir systems & 35 (0)\\
		HW Vir system with solved orbits & 17 (17)\\
		ellipsoidal deformation & 19 (11) \\
		Doppler beaming & 16 (1) \\
		\hline
	\end{tabular}
\end{table}

\begin{figure}
    \centering
    \includegraphics[width=\linewidth]{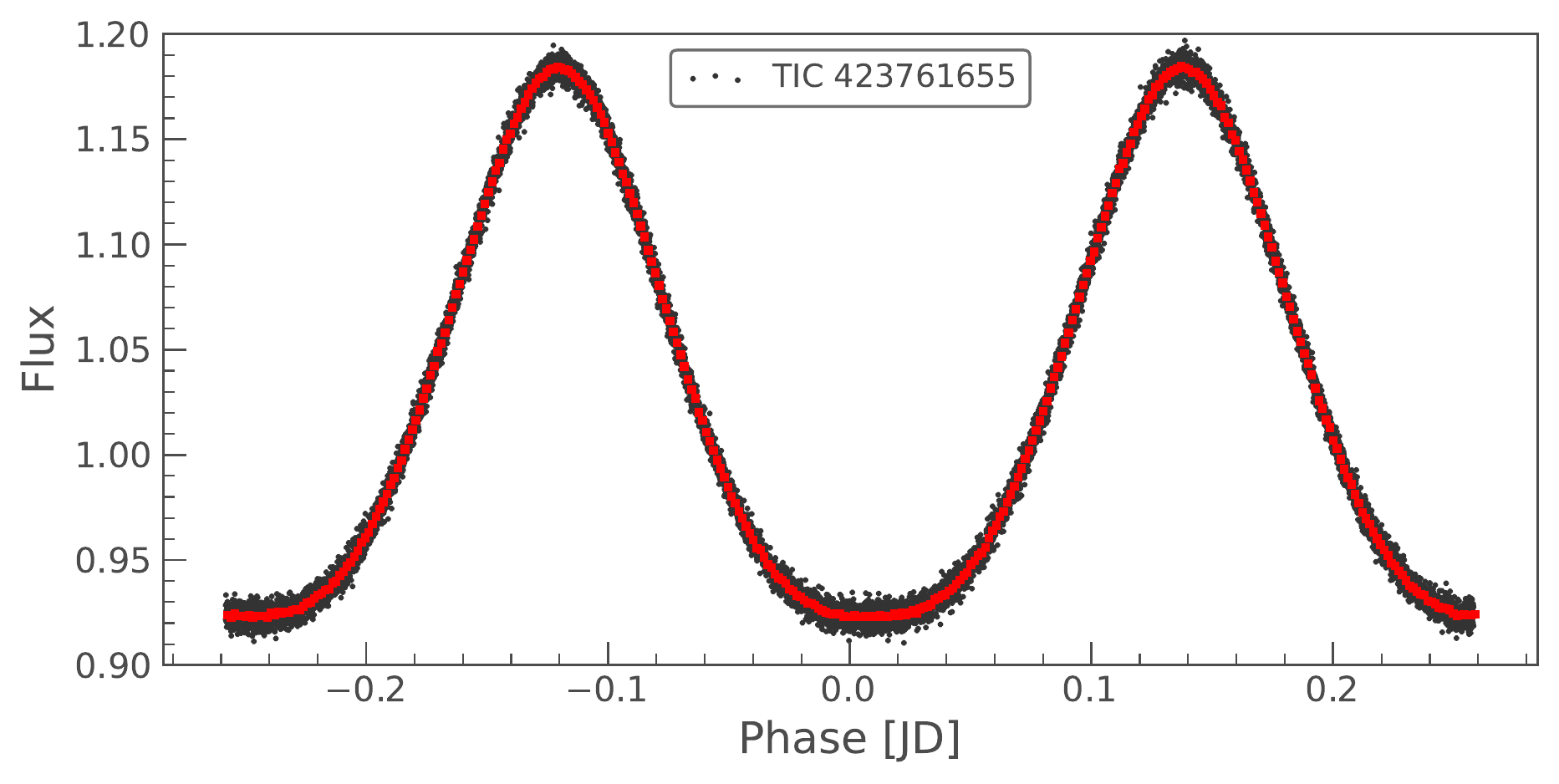}
    \caption{Example \textit{TESS} light curve of a reflection effect system (EC01578-1743). The light curve is shown phase-folded to the orbital period (black points) and also binned (red points).}
    \label{refl_example}
\end{figure}
\begin{figure}
    \centering
    \includegraphics[width=\linewidth]{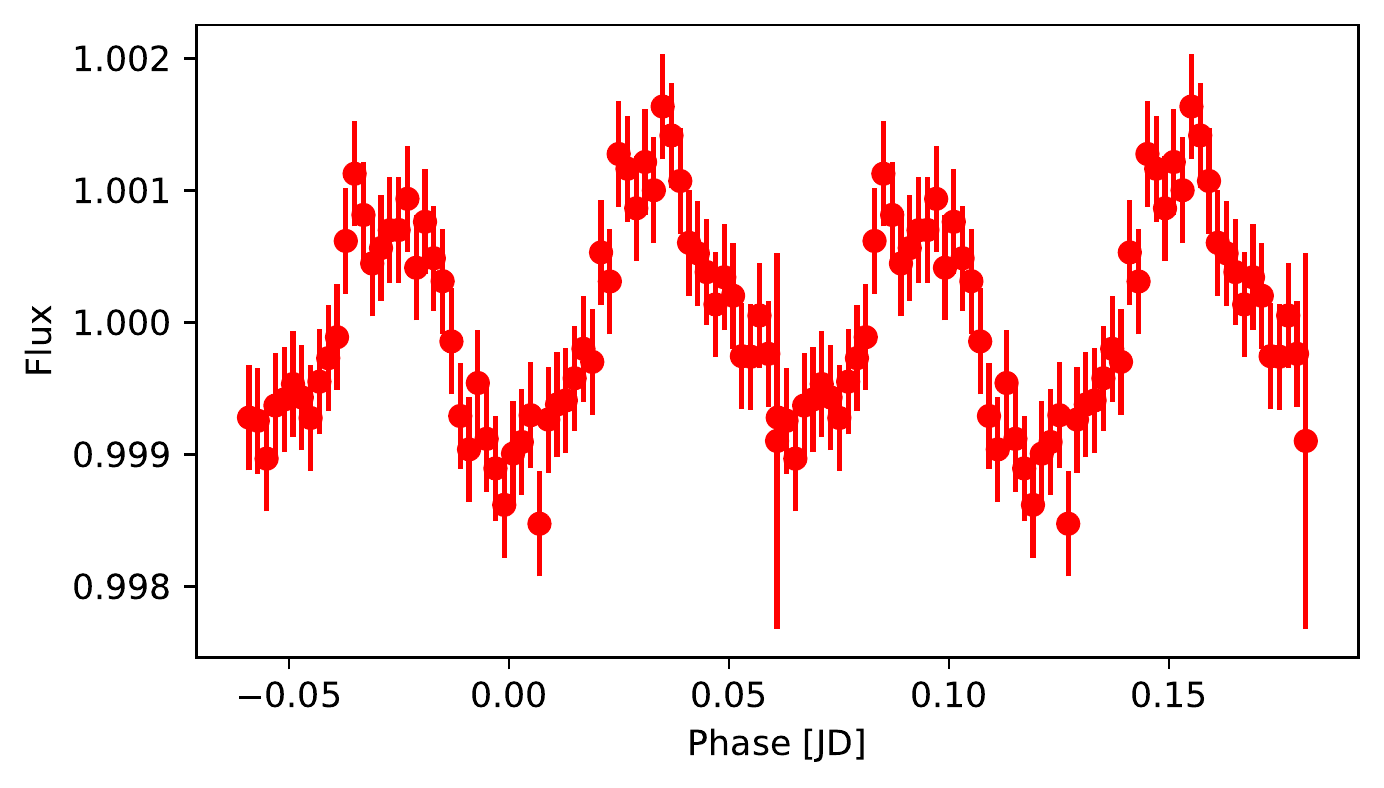}
    \caption{Example \textit{TESS} light curve of an ellipsoidal system (PG1043+760) showing additionally Doppler beaming. The light curve is shown phase-folded to the orbital period and binned.}
    \label{ell_example}
\end{figure}

To look for reflection effect systems in the \textit{TESS} light curves, we have searched \textit{TESS} sectors 1-36 for variability in all stars brighter than $G<16\,\rm mag$ from the \textit{Gaia} DR2 catalogue of hot subluminous stars \citep{geier_gaia_catalog}, as well as the catalogue of spectroscopically confirmed hot subdwarf stars (in total 2883  targets with 2-min cadence light curves and 353 targets with 20-sec cadence light curves).  \citep{geier:gaia}.% using a Lomb-Scargle periodogram \citep{lomb,scargle} derived with the python package \software{Astropy \citep{astropy:2013, astropy:2018}}. 
We have used the light curves made available by the \textit{TESS} Science Processing Operations Center (SPOC) through the Barbara A. Mikulski Archive for Space Telescopes MAST\footnote{\url{https://mast.stsci.edu/}}, using the PDCSAP flux, which corrects the simple aperture photometry (SAP) by removing instrumental trends, as well as contributions to the aperture expected to come from neighbouring stars other than the target of interest given a pre-search data conditioning (PDC). This is essential for \textit{TESS}, as the pixel size is almost $21\,\rm arc\,sec$. Through the CROWDSAP parameter, the pipeline also provides an estimate of how much of the flux in the aperture belongs to the target. To avoid possible zero-point inconsistencies between different sectors, we divided the flux by the mean flux in each sector for each star.

We used the Python package Astropy \citep{astropy:2013, astropy:2018} to calculate the Lomb-Scargle periodogram \citep{lomb,scargle} of all light curves up to the Nyquist frequency, oversampling by a factor of 10. Light curves were then phase-folded to the period corresponding to the strongest peak, or twice this period for ellipsoidal systems, which have first harmonic peaks stronger than the fundamental orbital frequency. Our custom script that downloads the light curves and generates diagnostic plots with the periodogram and phase-folded light curves is publicly available\footnote{\url{https://github.com/ipelisoli/TESS-LS}}. We visually inspected the diagnostic plots for all targets to confirm any variability and selected all objects showing a reflection effect (with and without eclipses), as well as stars showing ellipsoidal deformation. All targets with confirmed light variations can be found in Table \ref{refl}. 

Additionally, we inspected the \textit{TESS} or \textit{K2} light curves of all hot subdwarfs with with orbits characterized by radial velocity measurements \citep[][and references in Table \ref{tab}]{Kupfer2015}. All light curves were downloaded, phase-folded to the orbital period, and binned using the Python package \textsc{lightkurve} \citep[][]{lightkurve}\footnote{\url{https://docs.lightkurve.org}}. We computed the periodogram around the orbital period to search for any small peaks resulting from weak reflection or ellipsoidal deformation signals. For targets without any variations, we phase-folded the light curve to the orbital period derived by time-resolved spectroscopy and determined the signal-to-noise ratio. The results of our search are shown in Table \ref{search}, \ref{no_var}, and \ref{tab}. Example TESS light curves of a reflection effect system and an ellipsoidal system additional showing Doppler beaming can be found in Fig. \ref{refl_example} and \ref{ell_example}. The complete set of light curves, along with full details regarding our modeling and analysis methods, will be presented in an additional paper (Paper II, Schaffenroth et al. in prep.).

\section{Characterizing the primary star}
\subsection{Absolute magnitude and reduced proper motion}
\subsubsection{Method}
Both hot subdwarf and hot WD binaries containing a cool, low-mass companions can show a reflection effect \citep{erebos}, as can some sdOs that are central stars of planetary nebula (CSPN). In order to determine the true nature of the primary star, we used the colors, parallaxes, and proper motion from \textit{Gaia} EDR3 \citep{gaia_edr3}, as was done in \citet{erebos} for newly discovered HW Vir systems. Using the \emph{Gaia} $G$ magnitude together with the parallax, we could determine the absolute magnitude of our targets using the distance modulus ($G - M_G = 5 \log_{10} d - 5$). We ensured that all of our targets but one (which we identified as a potential triple system) had a small uncertainty in their parallax ($\lessapprox 10\%$) and a Renormalised Unit Weight Error (RUWE) below 1.4 \citep[e.g.,][]{penoyre22}. A higher RUWE indicates potential problems with the parallax. 

Another way to confirm our target selection is to determine the reduced proper motions $H_G = G + 5(\log \mu + 1)$. Stars that are farther away should show less transverse velocity on average than those that are closer, and the reduced proper motion is therefore a proxy for the distance; closer objects should have larger reduced proper motions. Typically, hot subdwarfs show reduced proper motions between 5 and 14 mag \citep[e.g.][]{erebos}.

\subsubsection{Results}
\begin{figure}
    \centering
    \includegraphics[width=\linewidth]{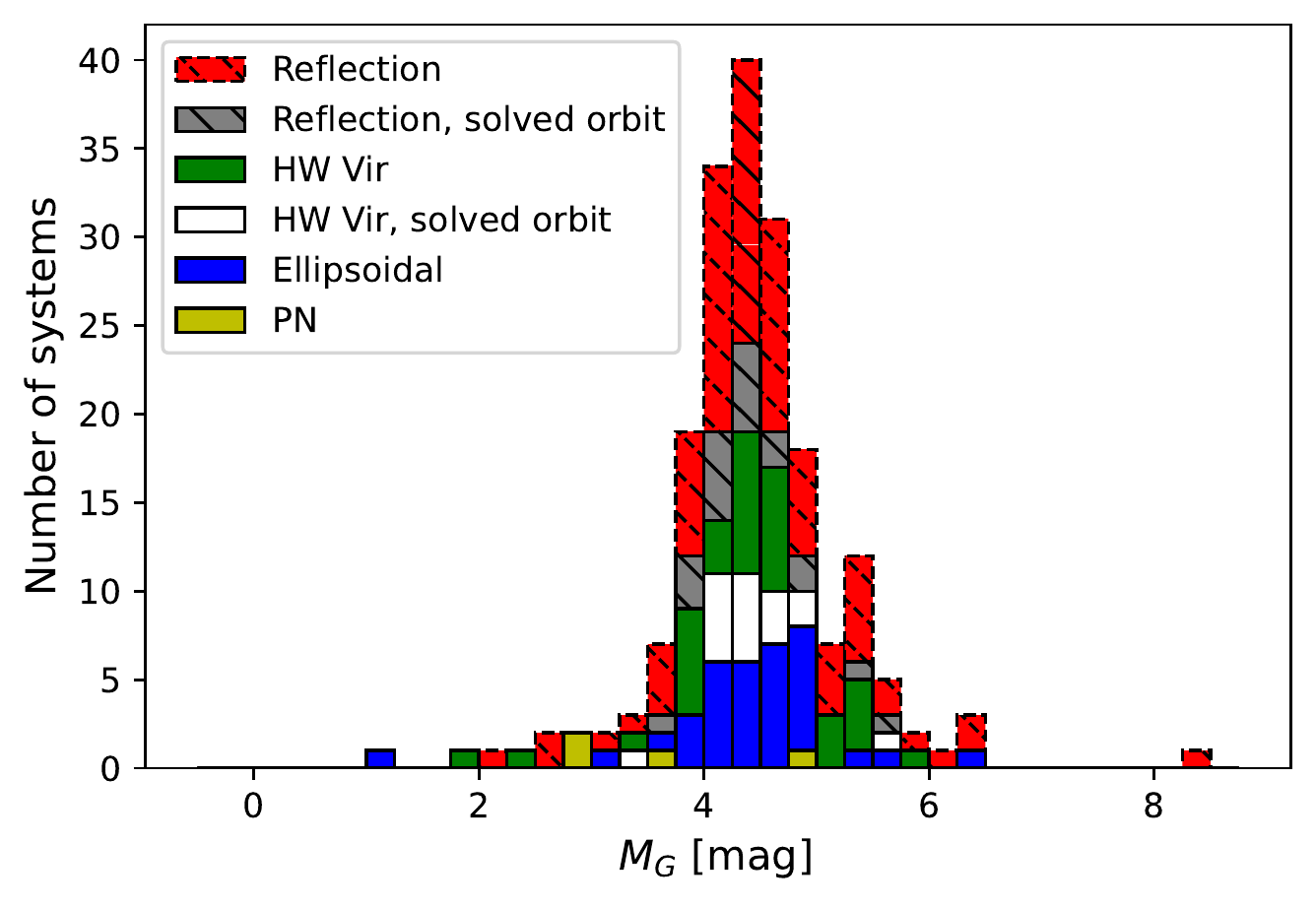}
    \caption{\emph{Gaia} absolute magnitude $M_G$ of all our targets divided into different groups according to the different light curve variations they show as shown in the legend.}
    \label{abs_mag}
\end{figure}

\begin{figure}
    \centering
    \includegraphics[width=\linewidth]{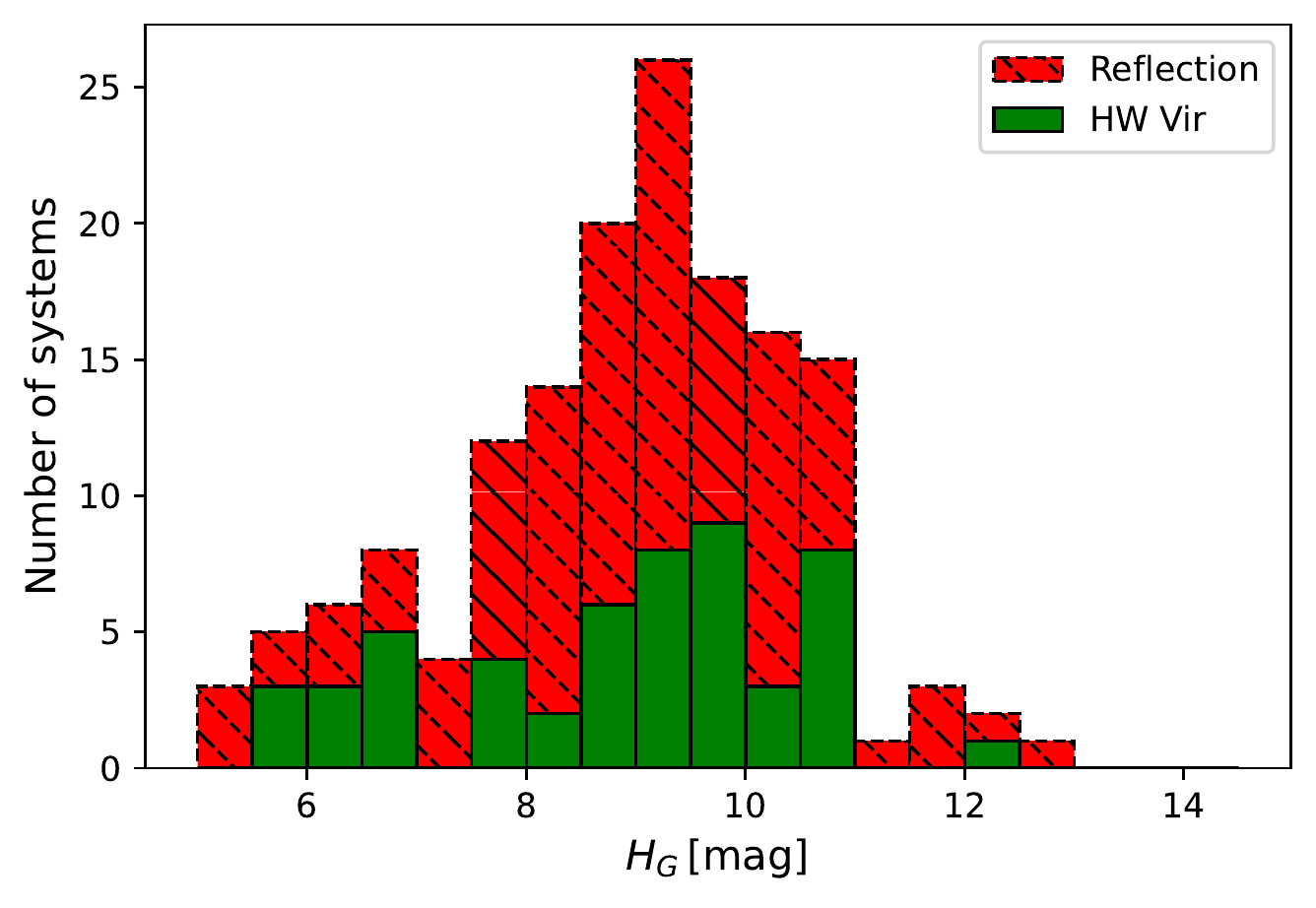}
    \caption{Reduced proper motion of all our reflection effect systems (eclipsing and non-eclipsing).}
    \label{red_pm}
\end{figure}

\begin{figure}
    \centering
    \includegraphics[width=\linewidth]{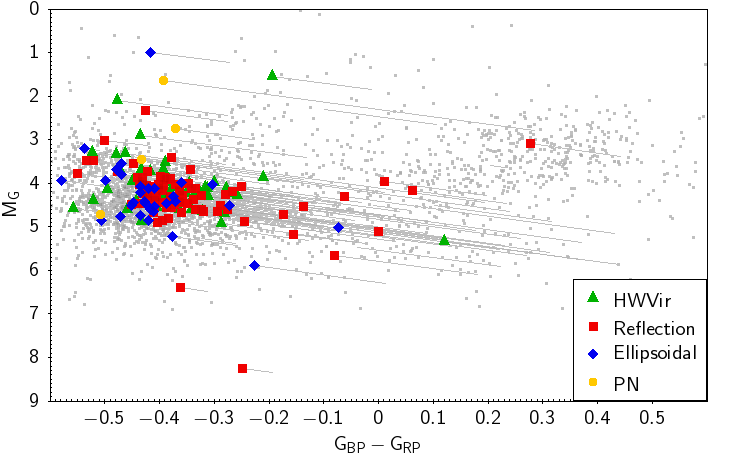}
    \caption{$G_{\rm BP}-G_{\rm RP}$ vs $M_G$ diagram. The targets are again grouped according to their light variations. All targets have been corrected for interstellar extinction using Stilism\protect\footnotemark. The correction is shown with the grey lines. In comparison the known sdO/Bs taken from \citet[][]{geier:gaia} are shown with the grey data points.}
    \label{bp-rp}
\end{figure}
\footnotetext{\url{https://stilism.obspm.fr/}}

\begin{figure}
    \centering
    \includegraphics[width=\linewidth]{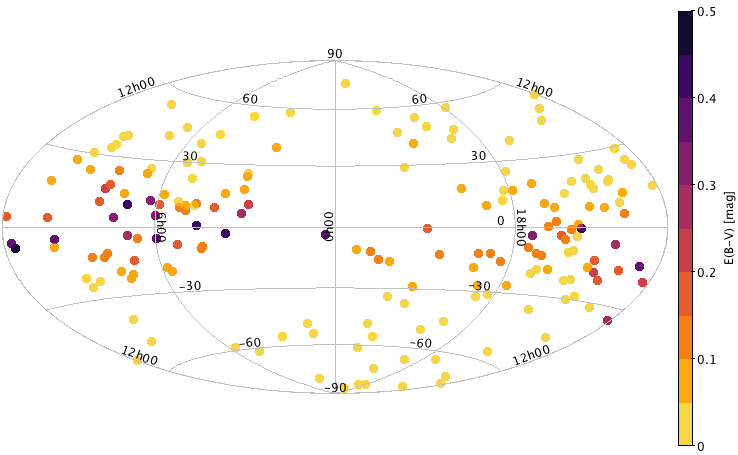}
    \caption{Position of our targets on the sky (in galactic coordinates). The color coding is giving by the color excess E(B-V).}
    \label{sky}
\end{figure}

\begin{figure}
    \centering
    \includegraphics[width=\linewidth]{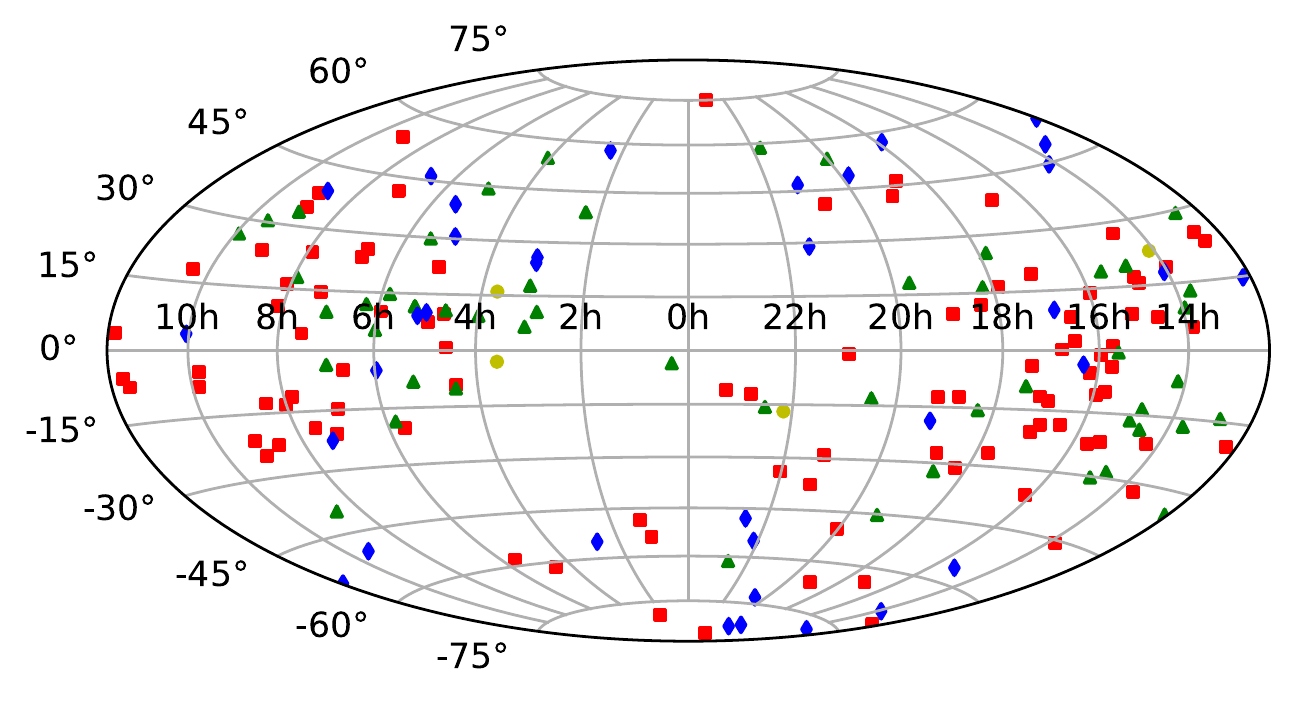}
    \caption{Comparison of the position of our targets on the sky (in galactic coordinates). The green triangles mark HW Virs, the red squares reflection effect systems and the blue diamonds ellipsoidal systems, and the yellow circles CSPNs.}
    \label{sky2}
\end{figure}

The results are found in Table \ref{refl}.
Inspecting the absolute magnitude $M_G$ distribution of all our targets (Fig. \ref{abs_mag}), we see that it peaks around $M_G=4.5$, as expected for hot subdwarf stars \citep{geier:gaia}. We have only one target with $M_G>7\,\rm mag$, which is most likely a WD primary. We also have some targets with $M_G<3\,\rm mag$, which are known CSPN, or pre-ELM WD. %The  %and some new ones in this region, also pre-ELM WD most likely. 

  Our reduced proper motion distribution shown in Fig. \ref{red_pm} also confirms that our targets are most likely hot subdwarf stars.

Since hot subdwarfs are of spectral type O and B, they have temperatures between 25\,000 to 50\,000 K and blue colors. Their luminosities are lower than main sequence stars and higher than hot white dwarfs. To check where we find our targets in the color-magnitude diagram, we plot a $G_{\rm BP}-G_{\rm RP}$ vs $M_G$ diagram (see Fig. \ref{bp-rp}). Both the absolute magnitude and $G_{\rm BP}-G_{\rm RP}$ color were corrected for interstellar extinction using 3D maps \citep{stilism, stilism2}. Our targets are located at $-0.5 <G_{\rm BP}-G_{\rm RP}<0.3$, with most of the targets clustering at $G_{\rm BP}-G_{\rm RP}<-0.25$. There is a slight trend that targets with $G_{\rm BP}-G_{\rm RP}>-0.25$ seem to have smaller $M_G$. As all of those targets show a high extinction, this trend can most likely be explained by insufficient correction of the interstellar extinction. The distribution of our targets on the sky (Fig. \ref{sky}) shows that most of the targets with high extinction are found close to the galactic plane, up to $\pm 20^\circ$ away. The comparison with the known sdO/Bs from \citet[][]{geier:gaia} shows quite a good agreement. One target is found with $G_{\rm BP}-G_{\rm RP}=0.3$ at an absolute magnitude $M_G=3$, consistent with known composite sdB stars. Only a few of our targets are found at $G_{\rm BP}-G_{\rm RP}<-0.45$, which is probably due to the fact that most of them are cooler helium-core burning sdB stars rather than evolved sdO stars. The comparison of the position of all our targets grouped together by the observed light variations (Fig. \ref{sky2}) shows that all different target types seem to be equally distributed on the sky.

\subsection{Spectral energy distribution}\label{sed_fit}

\begin{figure}
    \centering
    \includegraphics[width=\linewidth]{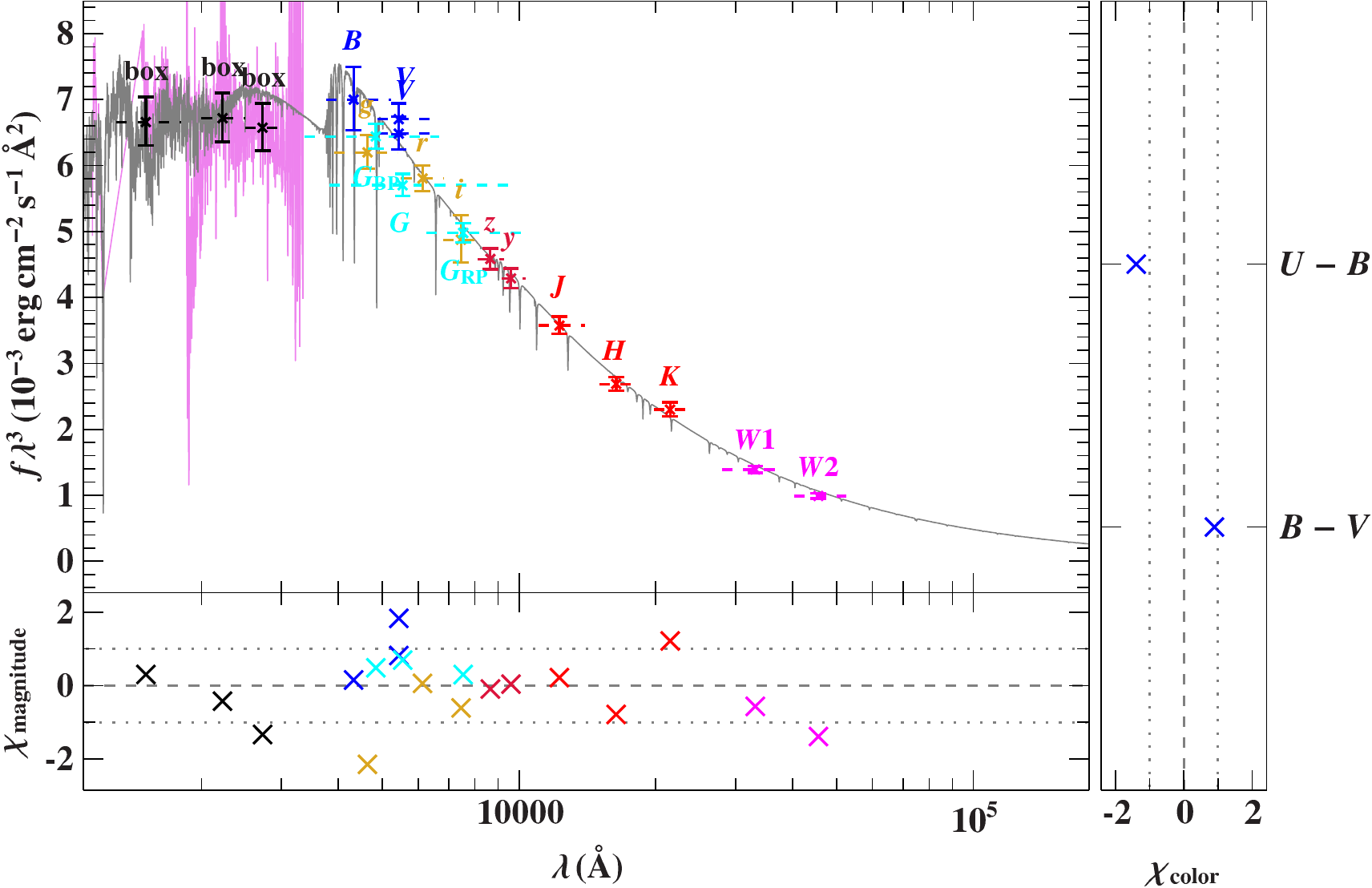}
    \caption{Example of an SED fit (for the sdB+WD system PG1519+640).}
    \label{sed_example}
\end{figure}

\subsubsection{Method}
To confirm a candidate's status as a hot subdwarf, we need to derive the effective temperature $T_{\rm eff}$ and surface gravity $\log g$. The best way to determine atmospheric parameters is to observe and model the target's spectrum. However, it is also possible to determine $T_{\rm eff}$, as well as the radius and the luminosity, by fitting the spectral energy distribution (SED) with synthetic spectra and combining this with the distance from the \textit{Gaia} parallax \citep[see][for more details on this method]{heber:2018,sed_andreas}. %; the rest will be presented with full analysis details in Paper II. 
The shape of the SED gives us $T_{\rm eff}$ as well as the interstellar reddening. And by comparing the observed and synthetic flux $f(\lambda)$ and $F(\lambda)$ respectively, we can derive the angular diameter $\theta=0.5\sqrt{\frac{f(\lambda)}{F(\lambda)}}$, which can be used to derive the radius $R=\theta/(2\varpi)$ and the luminosity $L/L_\odot=(R/R_\odot)^2(T_{\rm eff}/T_{\rm eff,\odot})^4$ by using the \emph{Gaia} parallax $\varpi$ and parallax offset.

Using the $\log g$ determined by the spectral fitting and the radius determined by the SED and \textit{Gaia} distance, we can also derive the mass of the hot subdwarf  $M=gR^2/G$ for the hot subdwarf binaries with known atmospheric parameters.

\subsubsection{Results}
One example SED fit is shown in Fig. \ref{sed_example}.
Unfortunately, the SED fitting is not straightforward for the reflection effect systems since our targets show light variations and the photometry we used from the literature was taken at a random phase. 

\begin{figure}
    \centering
    \includegraphics[width=\linewidth]{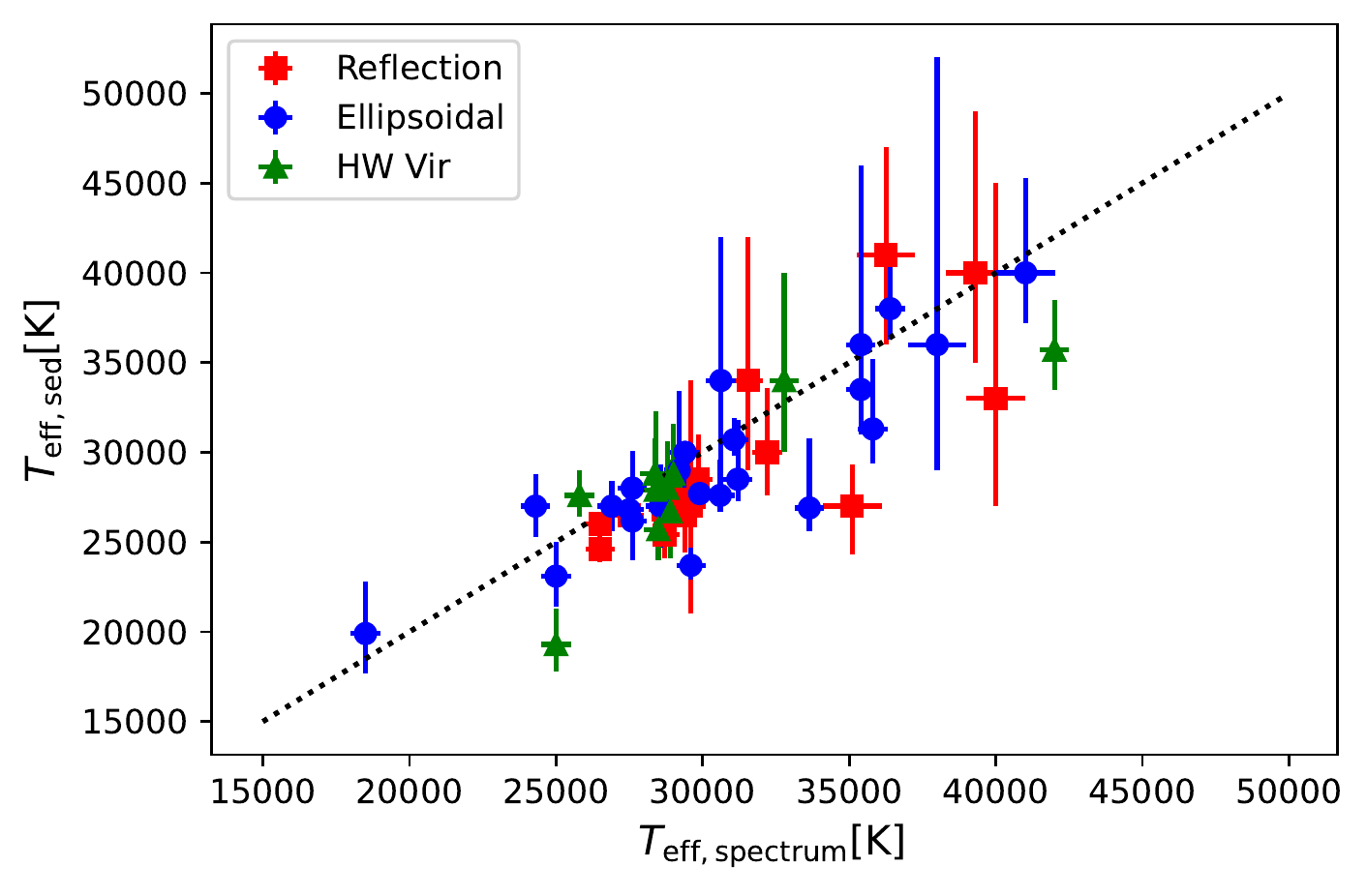}
    \caption{Comparison of the effective temperature determined by spectral fitting and spectral energy distribution fitting. Blue circles mark systems showing ellipsoidal deformation, green triangles mark HW Vir systems and red squares mark reflection effect systems. }
    \label{temp_sed}
\end{figure}

In light of the above, we tested our method on reflection effect and ellipsoidal systems with known atmospheric parameters and sufficient photometric data (see Table \ref{sed_known} for the results). The comparison between effective temperatures determined by spectral fitting and spectral energy distribution fitting (Fig. \ref{temp_sed}) shows that for systems with $T_{\rm eff}\lessapprox 32\,000\rm\, K$, the fitting of the SED can determine the $T_{\rm eff}$ very well if we neglect infrared photometry, since the contribution of the companion gets larger there. UV (IUE or Galex FUV or NUV) and SDSS u' photometry are essential for disentangling the effect of $T_{\rm eff}$ and interstellar reddening on the SED by covering the Balmer jump, so we exclude all targets without sufficient UV photometry from the SED fitting. For hotter systems, we see a larger scatter. For $T_{\rm eff}>42\,000\,\rm K$ the Balmer jump is not visible anymore, and so the temperature cannot be derived anymore without constraining the interstellar reddening. There is also a slight tendency that the SED fitting derives smaller temperatures than the spectral fitting.

\begin{figure}
    \centering
    \includegraphics[width=\linewidth]{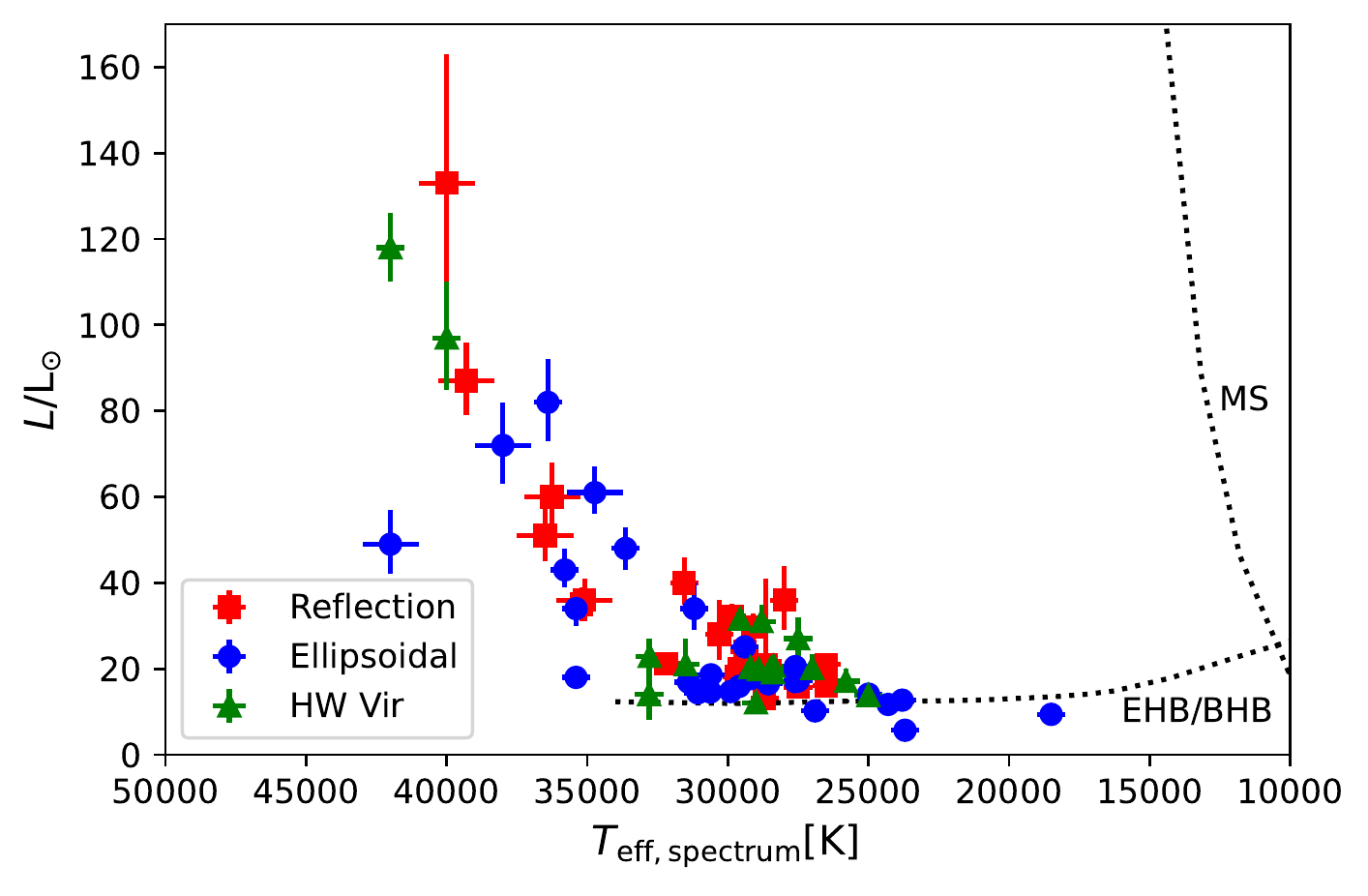}
    \caption{Hertzsprung-Russell diagram of our targets with known atmospheric parameters. Blue circles mark systems showing ellipsoidal deformation, green triangles mark HW Vir systems and red squares mark reflection effect systems. The dotted lines mark the zero-age main sequence or EHB/BHB.}
    \label{hrd_sed}
\end{figure}

Using the derived luminosities, we construct a Hertzsprung-Russell diagram for reflection effect and ellipsoidal systems with spectroscopic parameters for the first time. This is shown in Fig. \ref{hrd_sed}. The sdBs on the EHB with temperatures below 33000 K are found at similar luminosities between 15 and 40 $\rm L_\odot$. At larger temperatures the luminosity increases with the temperature but also a larger scatter is visible resulting from larger differences in the radii. \citet[][]{nemeth:2012} showed that the He abundances and the difference in He abundance increases with the temperature. The larger scatter of the radii and hence the luminosity at larger temperatures may be related to those He abundance differences.

\begin{figure}
    \centering
    \includegraphics[width=\linewidth]{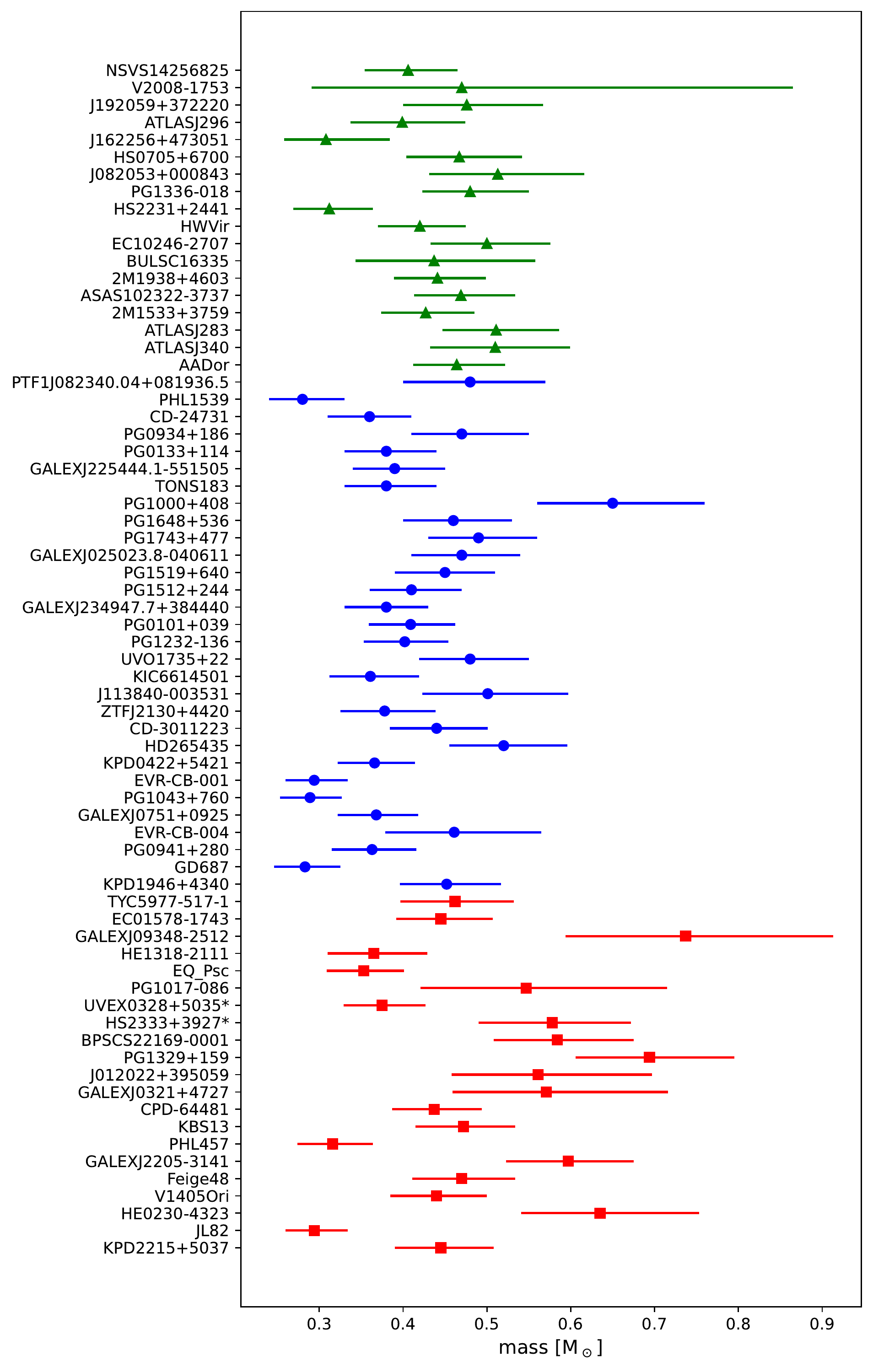}
    \caption{Masses determined by combining the spectroscopic analysis with the fit of the SED and the \textit{Gaia} parallaxes. Blue symbols mark systems showing ellipsoidal deformation, green symbols mark HW Vir systems and red symbols mark reflection effect systems. }
    \label{mass_sed}
\end{figure}

\begin{figure}
    \centering
    \includegraphics[width=\linewidth]{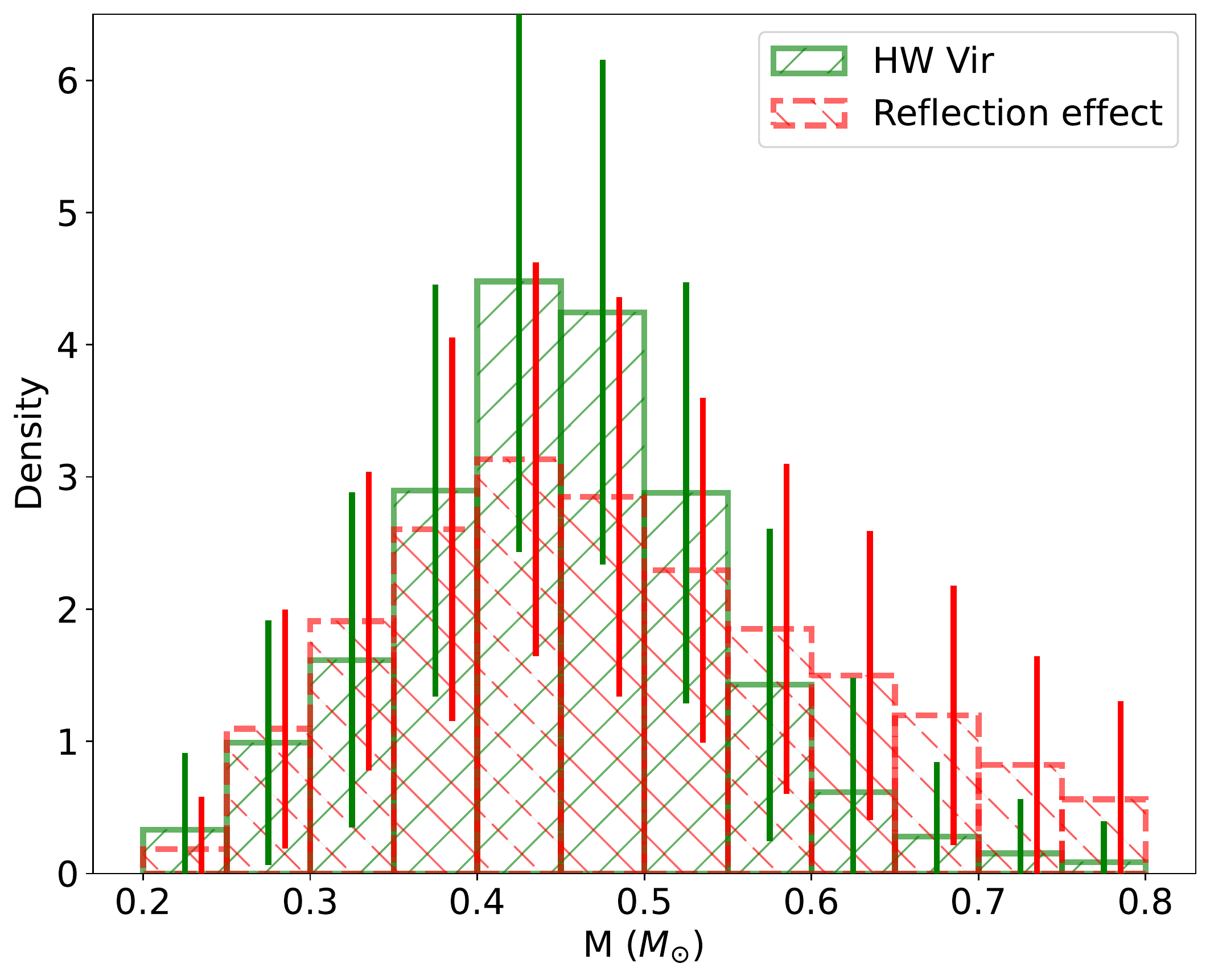}
    \includegraphics[width=\linewidth]{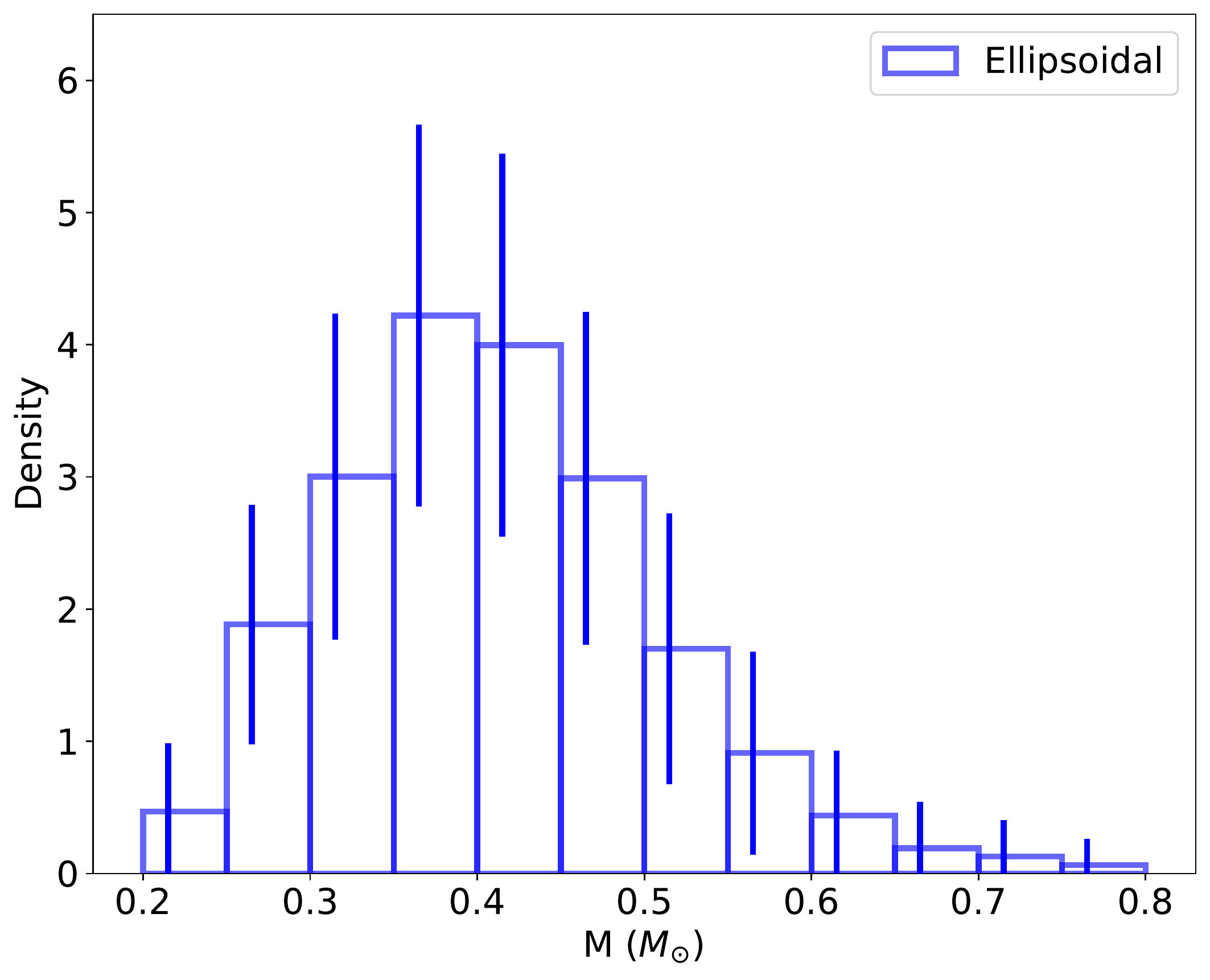}
    \caption{Histogram of the masses determined by combining the spectroscopic analysis with the fit of the SED and the \textit{Gaia} parallaxes.}
    \label{mass_hist_sed}
\end{figure}

\begin{figure}
    \centering
    \includegraphics[width=\linewidth]{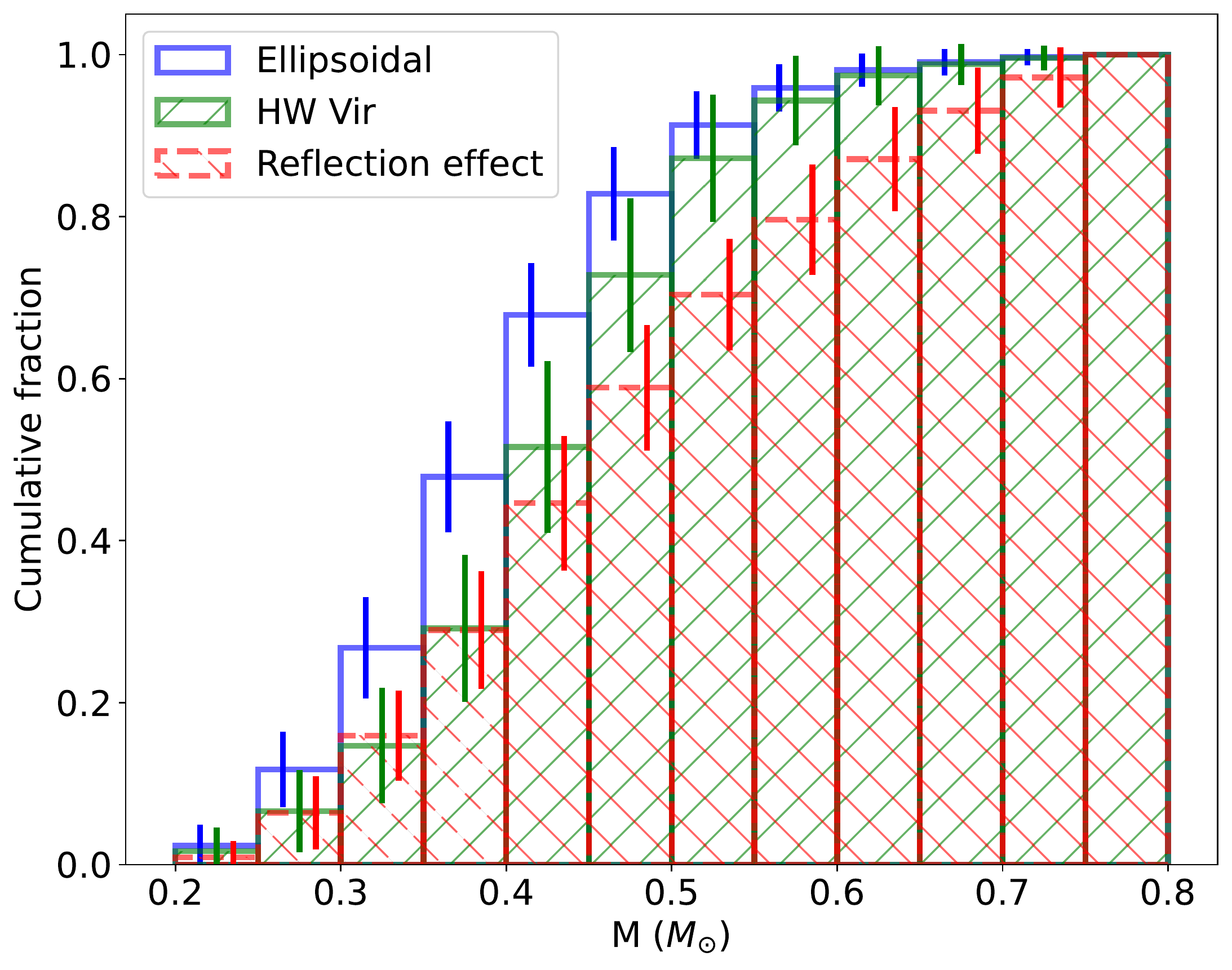}
    \caption{Cumulative distribution of the masses determined by combining the spectroscopic analysis with the fit of the SED and the \textit{Gaia} parallaxes.}
    \label{mass_hist_sed_cum}
\end{figure}

%The ellipsoidal systems show a distribution between 0.28 to 0.52 $\rm M_\odot$. The HW Vir systems show a larger peak around 0.47 $\rm M_\odot$  and a second small peak 0.3-0.35 $\rm M_\odot$. The reflection effect systems show a broader distribution between 0.29 and 0.74 $\rm M_\odot$ with a peak around 0.47 $\rm M_\odot$. 

With the spectroscopic $\log{g}$ and the radius from the SED fitting, we were able to derive the mass distribution of reflection effect, HW Vir and ellipsoidal systems. 
A similar approach was used in \citealt[][]{krzesinski}, which derived the mass distribution of pulsating hot subdwarf candidates with spectroscopic parameters to be a broad peak with the maximum at $0.45\,\rm M_\odot$. However, they state that their analysis might not be reliable and useful to derive masses of single systems.

Our result is shown in Fig. \ref{mass_sed}, \ref{mass_hist_sed} and \ref{mass_hist_sed_cum} and Table \ref{sed_known}.
For the HW Vir systems, we derive a mass of $0.46^{+0.08}_{-0.12}\,\rm M_\odot$ using a skewed normal distribution. As the typical mass error is about 0.05 $\rm M_\odot$ this suggests an intrinsically broader peak. The non-eclipsing reflection effect systems seem to have a broader peak with more higher-mass sdBs. As the only difference compared to the HW Vir systems is that they have no eclipses, we would not expect any difference. Determining atmospheric parameters from reflection effect systems has to be done with caution, as the contribution of the companion to the total flux changes with the orbital phase causing the reflection effect. So the atmospheric parameters have to be determined at or close to phase zero, when only the cool side of the companion is visible, or at the secondary eclipse, when the companion is occulted by the sdB in an eclipsing system. Most of the atmospheric parameters of the reflection effect systems have been determined from a single spectrum or co-added spectra at different orbital phases, causing systematic shifts to higher temperatures and higher $\log{g}$. This influences the determination of the radius and will result in a shifted mass. The HW Vir systems have been studied much more carefully, and so their determined atmospheric parameters are much more reliable.

The masses of the sdBs with white dwarf companions show a distribution with a similar width but a peak shifted to lower masses at $0.38^{+0.12}_{-0.08}\,\rm M_\odot$. The distribution also seems to be slightly asymmetric, extending to higher masses. The cumulative distribution shows the shift in mass more clearly, and shows that it is indeed significant.

The samples were taken from the literature and are not complete but are suffering from selection effects, which are not easy to determine. However, in the Gaia color-magnitude diagram (Figure. \ref{bp-rp}) and the sky distribution (Fig. \ref{sky2}), we can see that sdB+WD and the sdB+dM populations are overlapping well. The sdBs in the systems of both populations have been identified the same way by color selection. Hence, we expect that the selection effects should be similar for both populations and that they are comparable nevertheless.

The sdB+WD systems have been found preferably by RV variations in contrast to the HW Vir systems, as the sdB+WD systems show much smaller light variations. Both samples included only systems at the short period end of their period distribution (see Table \ref{refl}). A larger sample over a larger period range for both populations will be necessary to confirm our findings and also find or exclude differences of the sdB mass in systems with different orbital periods.

\begin{figure}
    \centering
    \includegraphics[width=\linewidth]{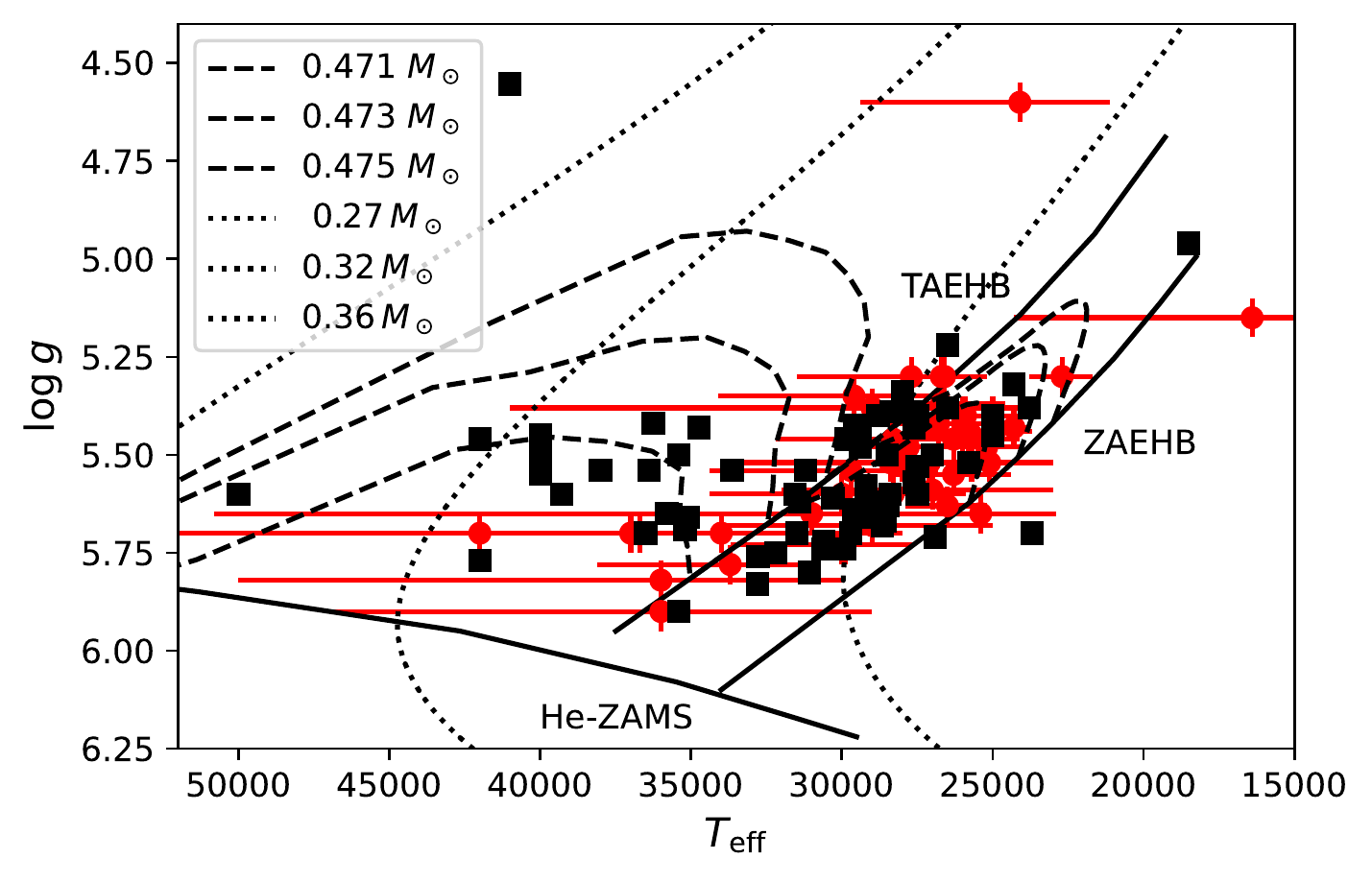}
    \caption{$T_{\rm eff}-\log{g}$ diagram of the sdB binaries with spectroscopic parameters (black squares) compared to the reflection effect candidates (red diamonds). The black solid lines mark the zero-age extreme horizontal branch (ZAEHB), the terminal-age extreme horizontal branch (TAEHB) and the He-main sequence (He ZAMS). The dashed lines are evolutionary tracks by \citet{Dorman:1993} for sdB masses of 0.471, 0.473, and 0.475 $\rm M_{\odot}$. The dotted lines are tracks for extremely low mass white dwarfs of a mass of 0.27, 0.32, and 0.36 $\rm M_{\odot}$ by \citet{althaus}.}
    \label{teff_logg__sed}
\end{figure}

By fitting the SED and combining this with the \textit{Gaia} parallax we can also constrain the atmospheric parameters of our reflection effect candidates, which do not have spectroscopic parameters, by fiting the SED and assuming a canonical mass for the sdB. From the radius we derive, we can constrain the $\log g$ in this way and constrain the atmospheric parameters for 44 targets with sufficient UV photometry. The results can be found in Table \ref{sed}. The atmospheric parameters are compared to the solved systems in the $T_{\rm eff}-\log{g}$ diagram (Fig. \ref{teff_logg__sed}). This shows that our reflection effect candidate systems are mostly found on the EHB.  Some of the candidates are on post-EHB tracks, which means the He in the core was exhausted and they are evolving away from the EHB. Only one candidate was found above the EHB, which could be a lower mass pre-He WD.

 %Most of the atmospheric parameters for the reflection effect systems have been derived by one spectrum at a random phase leading to a broader mass distribution. %Comparing the mass distribution if the reflection effect with or without eclipses to the ellipsoidal systems it can be seen that the there more sdBs in ellipsoidal systems that have lower masses. A reason could be that more of those are formed from young, higher mass systems and indeed many of the ellipsoidal systems are found towards the Galactic plane. 

%\begin{figure}
%    \centering
 %   \includegraphics[width=\linewidth]{logg_sed.pdf}
 %   \caption{Caption}
 %   \label{logg_sed}
%\end{figure}

%What we see is probably due t o the fact that for those the atmospheric parameters have been determined by a single spectra or co-added spectra at a random orbital phase over and the contribution of the companion to the total flux changes with the orbital phase causing the reflection effect. So the atmospheric parameters have to be determined at or close to phase 0 where only the cool side of the companion is visible or at the secondary eclipse, where the companion vanishes behind the sdB for an eclipsing system.

\section{The period distribution from light variations found by \textit{TESS}}

\begin{figure}
    \centering
    \includegraphics[width=\linewidth]{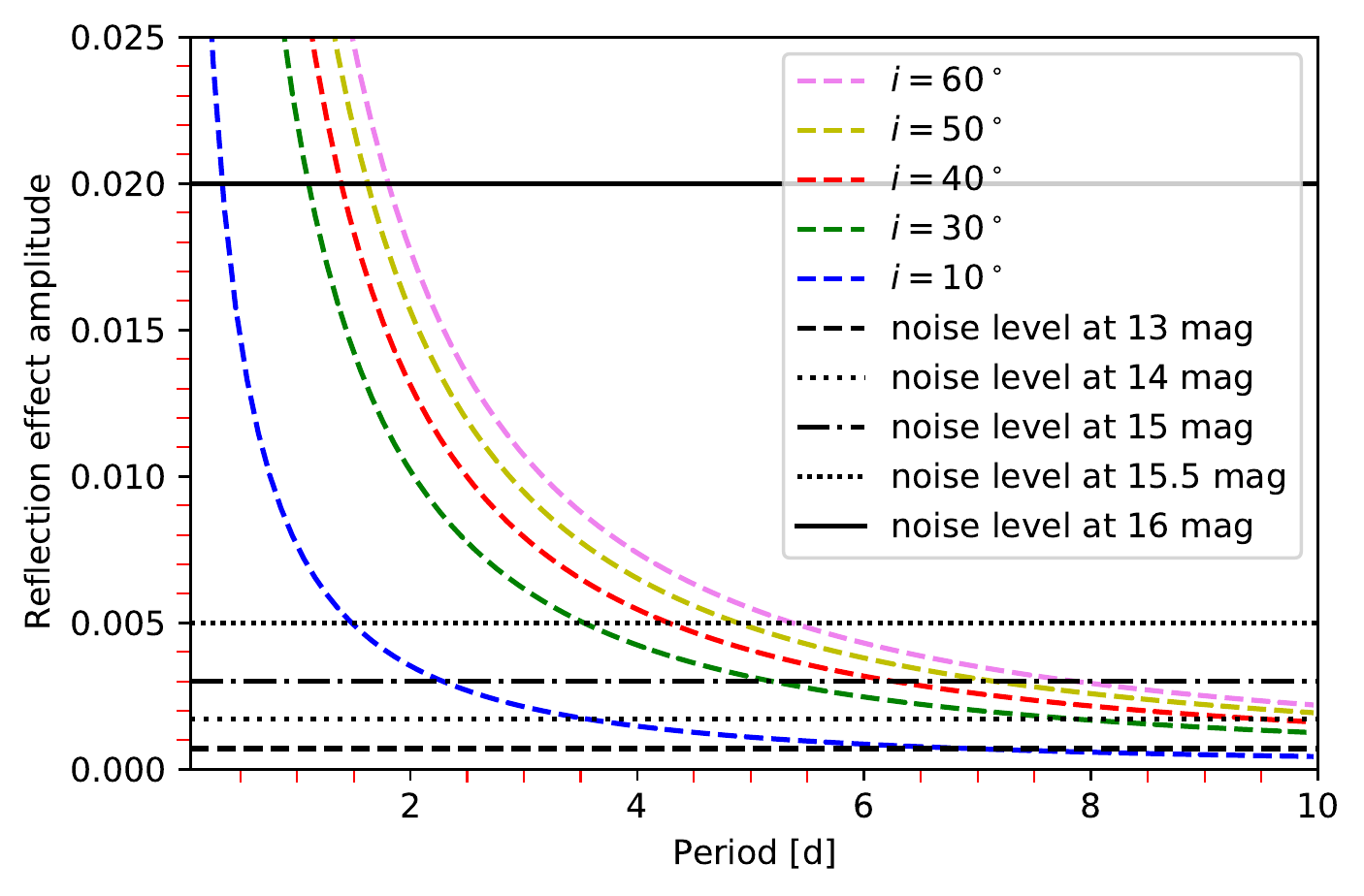}
    \caption{Amplitude of the reflection effect of a typical sdB+dM system for different periods and inclinations. The black lines mark the \textit{TESS} noise level for stars of different brightness.}
    \label{refl_ampl}
\end{figure}

\begin{figure}
    \centering
    \includegraphics[width=\linewidth]{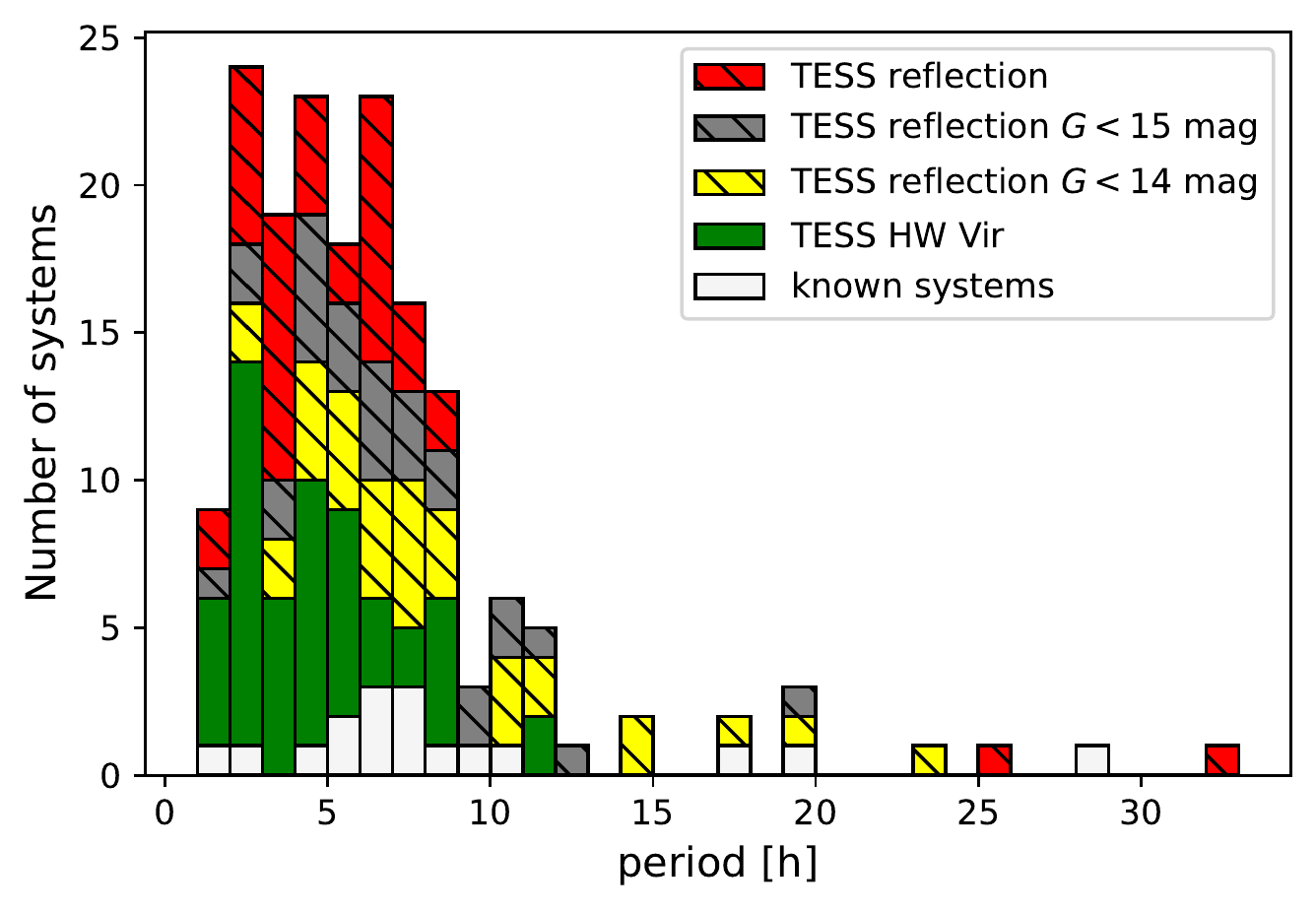}
    \caption{Period distribution of the reflection effect systems with and without eclipses observed by TESS. The known reflection effect systems are marked in white, the eclipsing reflection effect systems are marked in red, and the reflection effect systems without eclipses are marked in yellow, green and grey for systems with $G<13\,\rm mag, G<14\,mag$ and all other systems respectively.}
    \label{p_mag}
\end{figure}
\subsection{The selection effects of \textit{TESS}}
In order to judge the completeness of our reflection effect sample, we simulated the expected amplitude of the reflection effect for a typical sdB+dM system with different orbital periods and compared this to the noise level of the \textit{TESS} satellite for stars of different brightness (see Fig. \ref{refl_ampl}). This was done by checking the noise level in the light curves of different sdB stars of the same brightness not showing any variations in the light curve. For a 15 mag system, the detection limit is about 0.3\%. As expected, the amplitude of the variations decreases with lower inclination. But even with a low inclination of only $10^\circ$, we would expect to detect the reflection effect for a system brighter than 15 mag with \textit{TESS} up to two days, for higher inclinations of about $40^\circ$ up to about 6 days, and in inclinations of more than $60^\circ$ up to 8 days. Since the reflection effect becomes more sinusoidal at low inclinations, it is quite hard to distinguish it from other variations like pulsations or spots. Consequently, we will probably find low-inclination reflection effects only for the systems in which the period is already known from the RV curve. But the inclination should correlate with the period, and so this should not influence the period distribution we derive.  

Another selection effect could come from \textit{TESS} having such large pixels (21 arc-sec per pixel). If another star of comparable or higher brightness is close to the star, the light curve can become contaminated. %, as we have seen with
\textit{TESS} tries to correct for this additional flux through its reported PDCSAP flux, and it uses the CROWDSAP keyword in the header to quantify the contamination level. %Despite that for some targets it is visible that 
%the amplitude being a few percent different in different sectors in the case of HS2333+3927 showing that 
This correction can over- or underestimate the flux, and so the amplitude is not entirely reliable when a bright star is so close and it contributes significantly to the flux in the target pixel. This means that we might miss some reflection effect systems when unrelated stars are too close, but overall this correction seems to be quite good (a few percent difference; see Paper II for more details) and there is no reason why this should influence the period distribution of the detected systems.

As the amplitude of the ellipsoidal modulation is much smaller as can seen in Fig. \ref{ell_example}, this is very different for sdB+WD systems because we will only find the systems with the closest periods and/or highest mass companions by our light variation search, if the period is not known by RV variations for example.

\subsection{Period distribution of the reflection effect systems}

Taking all of this into account, we will never acquire a {\em complete} sample of reflection effect binaries from light curves alone, and the situation is even worse for the ellipsoidal systems. We do expect to find most reflection systems with higher inclinations observed by TESS up to periods around 7 days, as they can be identified from their light curve shapes with ease. Fig. \ref{p_mag} presents our observed orbital period distribution for sdBs with cool, low-mass companions. To ensure we do not see any difference with the brightness, we also checked the distributions of reflection effect systems of different brightness with a two-sample Kolmogorov–Smirnov test, but we could not find any significant differences. The period distribution shows that the reflection effect systems without eclipses tend to be found at periods longer than the eclipsing systems. This is expected as the eclipse probability decreases quickly with increasing separation distance and period. The period distribution shows a broad peak from 2-8 hours and falls off quickly on either end. There are very few systems with periods shorter than 2 hours, and none are below 1.2 hours. Above 8 hours, the distribution falls off quickly, and only a handful of systems are found beyond 20 hours. Despite our ability to detect systems with periods up to several days, the longest--period system we have found has a period of 35 hours.  Since we do not find any longer--period systems, they either do not exist, or they are incredibly rare. As \textit{TESS} continues to observe more and more reflection effect systems and increase the sample size, hopefully this question can be answered.

\section{The companions of the close sdB binaries with solved radial velocity curves}
\subsection{The nature of the companion}

As we have seen, the reflection effect generates a flux variation that is detectable at periods up to several days. The light variation from ellipsoidal deformation or Doppler beaming, however, is much weaker on average (below 0.1-0.2\%) at periods up to about one day and not detectable at longer periods. We can use these facts to differentiate between cool, low-mass companions and white dwarf companions (more details and the analysis of those systems is shown in Paper II) for the systems with periods known from RV variations.% showing much smaller often not-detectable variations in the light curves that do not show a significant variation

 We phased the available light curves of all hot subdwarfs with solved orbits (135 of the 165 systems have Kepler or \textit{TESS} light curves). Of those, 40 show a reflection effect and 33 show ellipsoidal deformation or Doppler beaming indicating that they have a white dwarf companion (see Paper II for more details). The rest do not show significant variations at the orbital period. We derived the signal-to-noise ratio for all light curves not exhibiting any variations. The result can be found in Table \ref{no_var}.

To constrain the nature of the companion, we used the amplitude estimates at a given period and inclination shown in Fig. \ref{refl_ampl}. Under the assumption all orbital plane orientations are equally probable, the probability of the inclination being lower than $10^\circ$ is only $1-\cos10^\circ=1.5\%$ \citep[][]{gray}. Therefore, we classify as sdB+WD systems all sdB binaries with light curves having a signal-to-noise ratio smaller than the amplitude expected for a reflection effect system observed at an inclination angle $<10^\circ$ (probability $>98.5\%$). 

Moreover, for companion masses larger than 0.45 $\rm M_\odot$ we would expect to see an infrared excess in the SED, if the companion would be a main sequence companion. Therefore, we also classify all sdB binaries with minimum companion masses $>0.45\,\rm M_\odot$ as sdB+WD system. The minimum companion masses can be derived by the mass function 
\begin{equation}
    f(M_1,M_2)=\frac{M_2^3\sin^3i}{(M_1+M_2)^2}=\frac{PK_1^3}{2\pi G},
\end{equation}
assuming a mass of 0.47 $\rm M_\odot$ for the sdB and an inclination of 90$^\circ$. We are unable to constrain the nature of the companion in this way for only 12 of our systems, as the noise in their light curves is too large and the minimum companion mass is too small. In total, this gives us 83 sdB+WD systems and allows us to constrain the nature of 75\% of all close sdB binaries with solved orbits.

Most of the sdB binaries with solved orbits have been detected by radial velocity variations, a method biased towards shorter periods, higher companion masses, and higher inclinations. Only very few of these were found by light variations. In this sample, about one-third of the sdB binaries have M dwarf or brown dwarf companions, and two-thirds have white dwarf companions. 

\subsection{The period distributions}

\begin{figure}
    \centering
        \includegraphics[width=\linewidth]{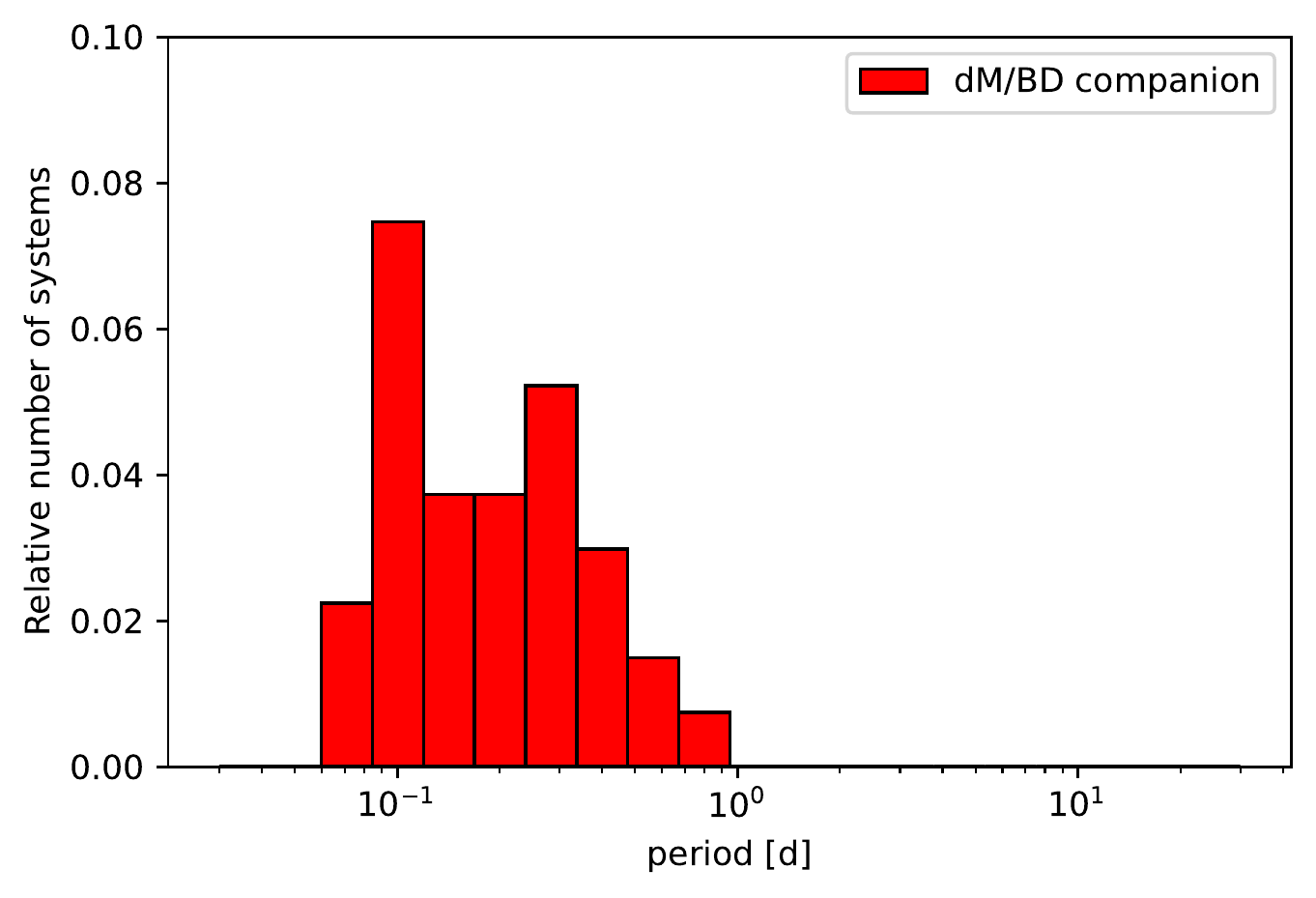}
        \includegraphics[width=\linewidth]{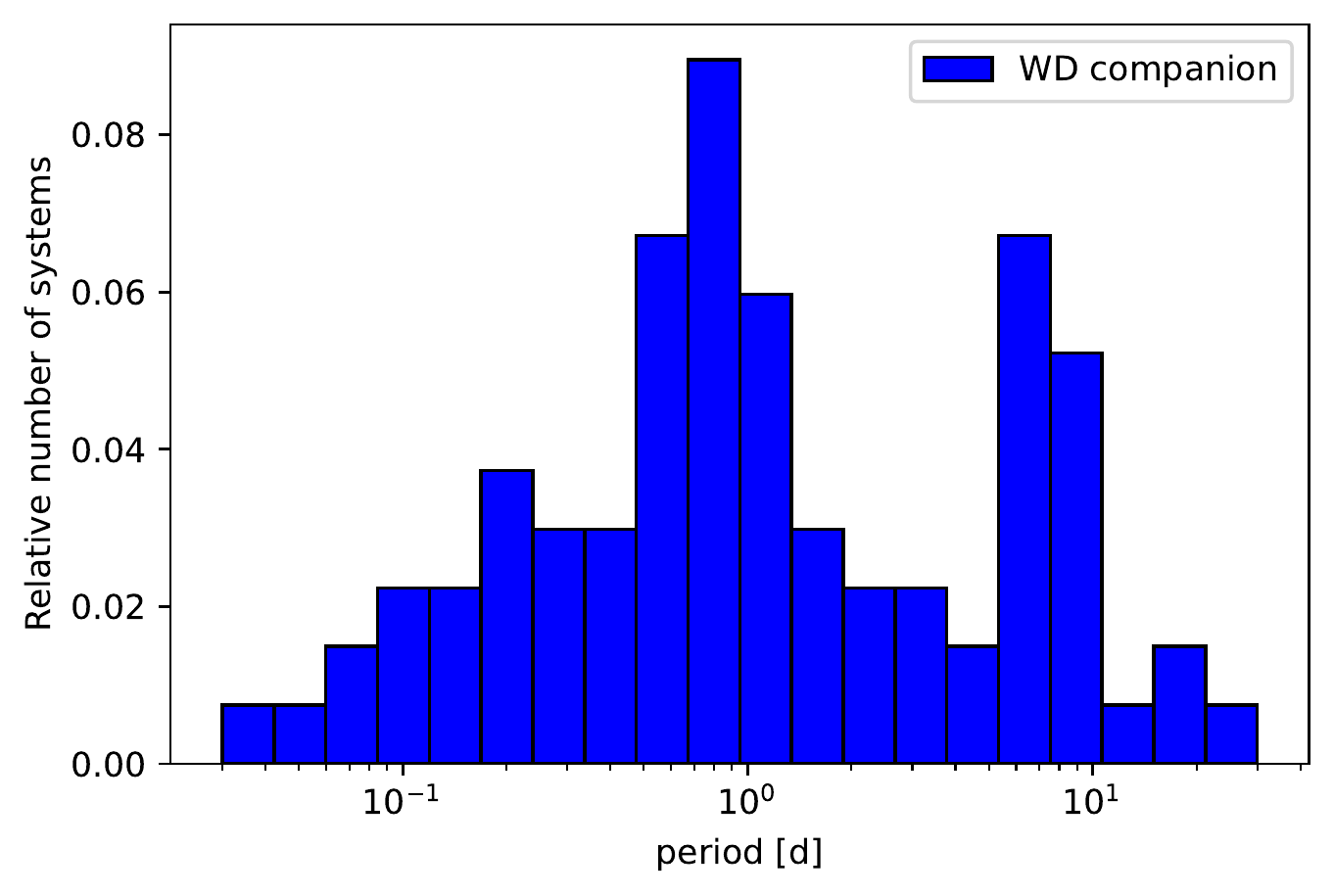}
        \includegraphics[width=\linewidth]{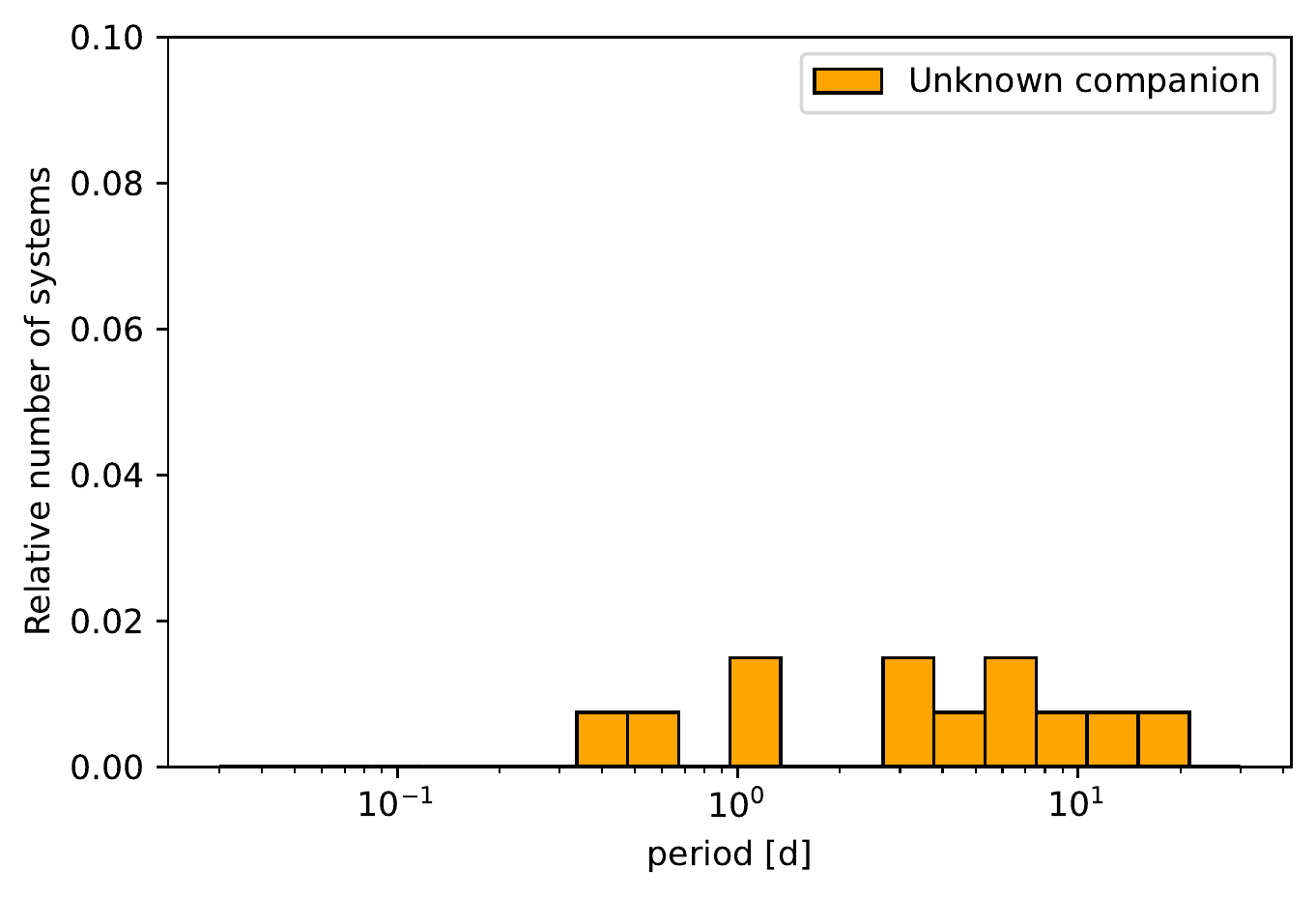}
    \caption{Period distribution of the hot subdwarf binaries with solved orbits with dM/BD companions in the top panel, with WD companions in the middle and unknown companions in the lower panel.}
    \label{period_dist}
\end{figure}

The updated period distributions of the dM/BD and WD companions and their differences are also interesting, as shown in Fig. \ref{period_dist}. We already discussed the distribution of the reflection effect and HW Vir systems showing periods from 2 hours to about 1 day. The systems with WD companions, on the other hand, show a broad distribution from just about one hour to 27 d. On top of this broad distribution we find two distinct peaks at around one day and around 5-10 days. The companion is still undefined only for a small number of systems. Most of them have periods longer than one day, agreeing well with the distribution of the WD companions, so it is likely they are also sdB+WD systems.

\subsection{The minimum companion masses}

\begin{figure}
    \centering
    \includegraphics[width=\linewidth]{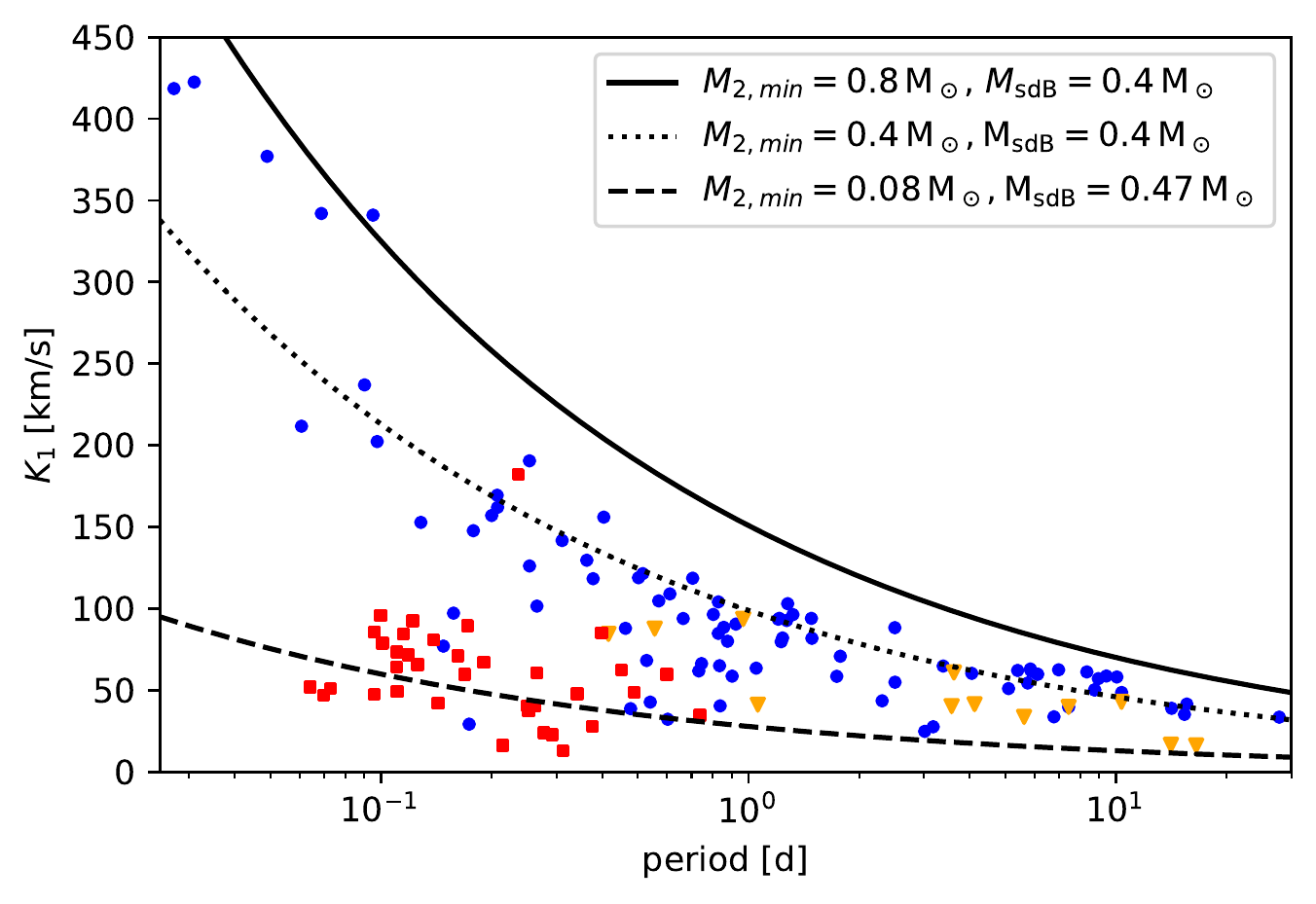}
    \caption{RV semi-amplitudes of all known short-period sdB binaries with spectroscopic solutions and with \textit{TESS} or Kepler light curves plotted against their orbital periods (red squares: dM/BD companions, blue circles: WD companions, yellow diamonds: unknown type). The lines mark a certain minimum companion mass derived
from the binary mass function (assuming 0.47 or 0.4 $\,M_\odot$ for the sdBs).}
    \label{rv_all}
\end{figure}

\begin{figure}
    \centering
    \includegraphics[width=\linewidth]{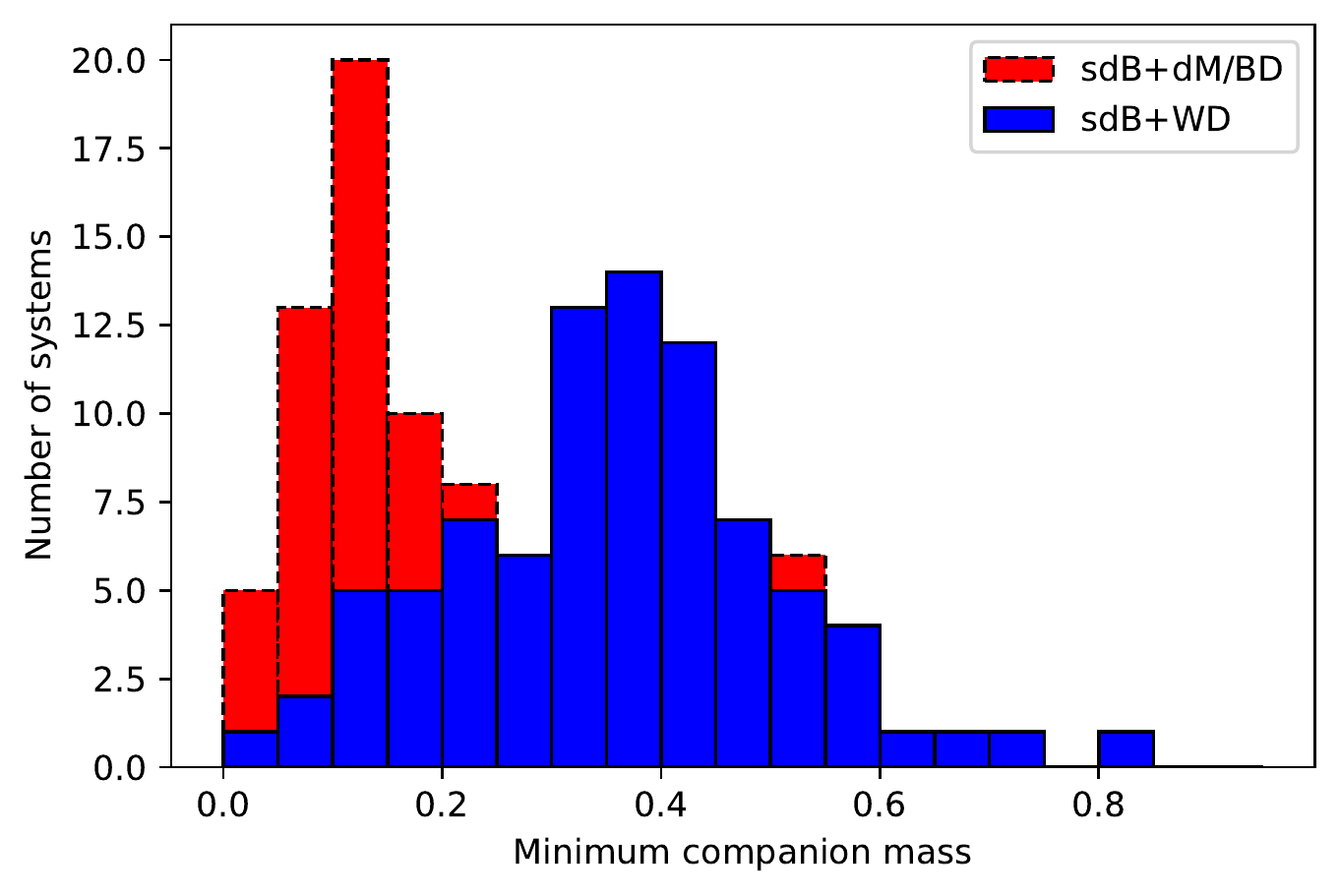}
    \caption{Minimum companion mass distribution of all known short-period sdB binaries with spectroscopic solutions and with \textit{TESS} or \textit{K2} light curves (assuming 0.47 $\,M_\odot$ for a sdB with dM/BD companion or 0.4 $\,M_\odot$ for a sdB with WD companion).}
    \label{min_mass}
\end{figure}

To get a clearer picture of the masses of the close companions to hot subdwarf stars, we update the plot of RV semi-amplitude versus orbital period for all known sdB binaries (as shown in \citealt[][]{Kupfer2015}) with spectroscopic solutions and with \textit{TESS} or Kepler light curves (Fig. \ref{rv_all}) as well as plot the minimum companion mass distribution in Fig. \ref{min_mass}.

Our new sample adds many more systems with companion mass constraints to this plot. As we have seen, it is possible to constrain the minimum mass of the companion from the RV semi-amplitude of the sdB and the orbital period, when assuming a mass for the sdB. This is given by the black lines for different periods. For a random distribution of system angles, the probability to have a system with inclination $>60^\circ$ is the same as for an inclination of $<60^\circ$, and so about half of the companions should have masses of only up to 20\% higher than the minimum companion mass. 

We find that systems with cool, low-mass companions cluster around the hydrogen-burning limit with masses up to $0.25\,\rm M_\odot$ with one exception. The white dwarf companions to sdB stars have higher minimum masses, and it looks like there are three different populations. At the shortest periods from approximately one to about three hours, a small group of WD companions with minimum companion masses around 0.7 to 0.8 $\rm M_\odot$ are found. %indicating a CO or ONeMg WD. 
Most of the WD companions are found in binaries with longer periods. Up to a period of about 4 days they seem to have significantly lower minimum companion masses with a mass ratio close to one (around $0.4\,\rm M_\odot$, when assuming an sdB mass of $0.4\,\rm M_\odot$). % indicating that many of the WD companions could be He WDs.
Systems with periods belonging to the second peak in the period distribution around 5-10 days show some indication of slightly higher minimum companion masses above $0.4\,\rm M_\odot$ up to $0.6\,\rm M_\odot$.

\section{Discussion and Conclusions}
Our light variation search increases the number of known reflection effect systems from 19 to 104 systems. Moreover, we detected 23 new sdB+WD systems showing tiny variations with amplitudes below $\sim 0.1\%$, due to Doppler beaming or ellipsoidal deformation in their light curves.

The characterization of the reflection effect systems in our sample shows that all except one have hot subdwarf primaries. The one exception was a system with a white dwarf primary. Similar results were found in other surveys, such as EREBOS \citep[][]{erebos}.
%The characterization of the primary star of our reflection effect systems shows that it is a hot subdwarf star with one exception, which is a white dwarf primary
%even though white dwarfs are much more common. 
The detection of a white dwarf primary is not a complete surprise since we selected targets from the \textit{Gaia} DR2 catalogue of hot subluminous stars, which does have some overlap with the white dwarf catalogue. Nonetheless, most of the primaries in our systems should be sdO/B stars. The reflection effect is only visible in hot white dwarfs, which are much rarer. Moreover, white dwarfs are much fainter than hot subdwarf stars. And since sdBs are mainly formed by binary evolution, the binarity rate of sdBs is much higher than of WDs. That is why reflection effect systems with hot subdwarf primaries will dominate all surveys for reflection effect systems.

To check the mass determination of the sdB using the SED and the \textit{Gaia} parallaxes, we compared the masses derived by this method with the masses derived by the light curve analysis of several ellipsoidal systems. This is shown in Table \ref{mass_sdb_wd}. The masses derived by the two different methods agree very well within the errors for all systems, thereby showing the validity of our spectrophotometric \textit{Gaia} distance method.

The comparison of the mass distribution of the sdB+dM and the sdB+WD (Fig. \ref{mass_hist_sed}) shows that they differ significantly. The mass distribution of sdBs with WD companions is shifted to lower masses compared to sdBs with dM/BD companions. This implies that sdBs with dM/BD companions come from a different population than sdBs with WD companions. The sdB+dM systems show a peak around the canonical mass for He burning and a few systems at higher and lower mass, as predicted by binary population synthesis models \citep[][]{han:2002,han:2003}. Those non-canonical systems can originate from young, higher-mass systems igniting He in the core under non-degenerate conditions or be pre-He WDs not massive enough for He-burning that are passing through the sdB region in the HRD. The sdB+WDs, on the other hand, show many more low-mass systems. The sdB binaries with massive companions are observed towards the Galactic plane, where younger stars are found. This indicates that those systems are preferably formed in younger populations than the sdB+dM stars. The other sdB+WD systems seem to be equally distributed on the sky (see Fig. \ref{sky2}).

\begin{table}
    \caption{Masses of the solved sdB+WD systems derived by light curve analysis and SED fitting}
    \label{mass_sdb_wd}
     \setlength{\tabcolsep}{1.5pt}
\begin{tabular}{llll}\hline\hline
    target & $M_{\rm sdB,SED}$ &$M_{\rm sdB,lc}$ & references\\
    & [$\rm M_\odot$] & [$\rm M_\odot$]&\\\hline
    KPD1946+4340 & $0.452^{+0.065}_{-0.056}$ & $0.47^{+0.03}_{-0.03}$&\citet[][]{bloemen11}\\
    CD-3011223   &  $0.44^{+0.061}_{-0.056}$&$0.47^{+0.03}_{-0.03}$&\citet[][]{cd-30}\\
    PTF1J0823+0819 & $0.48^{+0.09}_{-0.08}$ & $0.45^{+0.09}_{-0.07}$ & \citet[][]{kupfer:2017}\\
    EVR-CB-001 & $0.294^{+0.04}_{-0.034}$ &$0.21^{+0.05}_{-0.05}$ & \citet[][]{evr01}\\
    EVR-CB-004 & $0.461^{+0.104}_{-0.082}$ & $0.52^{+0.04}_{-0.04}$ & \citet[][]{evr04}\\
    ZTFJ2130+4420 & $0.378^{+0.061}_{-0.053}$ & $0.337^{+0.015}_{-0.015}$&\citet[][]{kupfer20}\\
    HD265435 & $0.59^{+0.17}_{-0.14}$ & $0.64^{+0.10}_{-0.09}$ & \citet[][]{pelisoli21}\\
    \hline
\end{tabular}
\end{table}

%\begin{figure}
%    \centering
%    \includegraphics[width=\linewidth]{m_r_TESS.pdf}
%    \caption{Mass-radius relation of the companions in the analyzed reflection effect systems compared to theoretical calculations by \citet[][]{baraffe:15}.}
%    \label{m-r}
%\end{figure}
%\begin{figure}
%    \centering
%    \includegraphics[width=\linewidth]{incl_distr.pdf}
%    \caption{Inclination distribution of the analyzed reflection effect systems. The black line shows the number of systems we expect, when we assume that the orientation of a sdB binary is uniformly distributed. Due to the projection effect it is much more
%likely to find binary systems at high rather than low inclinations.}
%    \label{incl}
%\end{figure}

The observation of space-based, high S/N light curves covering a time span of at least 27 days up to several months of so many sdBs gave us for the first time a large sample of reflection effect binaries. Since they were selected mainly from the \textit{Gaia} hot subdwarf catalogue and had no prior RV measurements, this gives us the first period distribution of sdB+dM systems selected {\em only} by light variations. The orbital period distribution of post-CE binaries is mainly dependent on the criterion for the ejection
of the CE \citep[][]{han:2002}, and so this distribution can be used to constrain the common-envelope phase when combined with the companion mass distribution as done in \citet[][]{ge22} for the sample of hot subdwarf binaries from \citet[][]{Kupfer2015} or comparing it to a modelled sdB binary sample using binary population synthesis. 

Aided by high-quality \textit{TESS} light curves, we could constrain the nature of the companion in 75\% of the sdB binaries with solved orbits and compare them. As seen in Fig. \ref{period_dist}, the period distribution of the sdB+dM systems is concentrated in a much smaller period range compared to the sdB+WD systems, which are found over a wide range of periods from 0.03-30 days. The distribution of the minimum companion masses found at a certain orbital period (Fig. \ref{rv_all}) shows that the companions in the reflection effect systems have minimum masses typical for BD/dM systems ($0.05-0.2\,\rm M_\odot$). There is no change with the orbital period visible. For the sdB+WD systems this is different. There seem to be two distinct groups of companion masses. At the shortest periods below 0.1 d, WD companions with high minimum masses around $0.8\,\rm M_\odot$ are found, which could be CO or ONe WDs. At longer periods, the WD companions seem to have significantly lower minimum masses, with masses around $0.4\,\rm M_\odot$. Many of those could be He-WD companions. At the longest periods the masses seem to be slightly higher indicating a third population of low-mass CO WD companions. This could suggest that sdB+WD systems at the shortest periods come from a different population with higher-mass progenitors having higher-mass companions than the longer period sdB+WD systems, which is consistent with predictions by binary population synthesis \citep[][]{han:2002,han:2003}.

The high signal-to-noise light curves allowed us to derive parameters for a large number of sdB+dM/BD and sdB+WD systems. Details of this light curve modeling and analysis are discussed in a separate paper (Paper II).

As \textit{TESS} continues to observe, the number of high-quality reflection effect and sdB+WD light curves will continue to grow. This will further increase our sample size and improve constraints on the mass and period distributions. Future spectroscopic and photometric surveys like 4MOST, BlackgGem, and Vera Rubin Observatory will also increase our sample size and our knowledge about these systems. %This allows us to study the common envelope phase, the sdBs and the companions to the sdB stars.

\begin{acknowledgements}
     This research made use of Lightkurve, a Python package for Kepler and \textit{TESS} data analysis \citep[][]{lightkurve}.
     
    % Based on observations collected at the European Organisation for Astronomical Research in the Southern Hemisphere under ESO programme(s) 080.D-0685(A), 082.D-0649(A), 092.D-0040(A).
     
     This paper includes data collected by the \textit{TESS} mission, which are publicly available from the Mikulski Archive for Space Telescopes (MAST). Funding for the \textit{TESS} mission is provided by NASA's Science Mission directorate.

     VS and SG acknowledge funding from the German Academic Exchange Service (DAAD PPP USA 57444366) for this
     project and would like to thank the host institution Texas Tech University for the hospitality. VS was funded by the Deutsche Forschungsgemeinschaft under grant GE2506/9-1. IP was partially funded by the Deutsche Forschungsgemeinschaft under grant GE2506/12-1 and by the UK’s Science and Technology Facilities Council (STFC), grant ST/T000406/1.

     TK acknowledges support from the National Science Foundation through grant AST \#2107982, from NASA through grant 80NSSC22K0338 and from STScI through grant HST-GO-16659.002-A.
     
     %IP acknowledges support from the UK's Science and Technology Facilities Council (STFC), grant ST/T000406/1. 
     BNB was supported by the National Science Foundation grant AST-1812874.
     
     We thank Uli Heber for comments on the manuscript. %and Lars Möller for sharing his radial velocity measurements with us. 
     We thank Andreas Irrgang for the development of the SED fitting tool and making it available to us.% We thank Alfred Tillich for observing some of the spectra used in this paper.
    % We thank Stephen Walser for helping with some of the SOAR and Chiron observations used in this paper. 
\end{acknowledgements}

\bibliography{biblio}

\begin{thebibliography}{105}
\expandafter\ifx\csname natexlab\endcsname\relax\def\natexlab#1{#1}\fi

\bibitem[{{Af{\c{s}}ar} \& {Ibano{\v{g}}lu}(2008)}]{PN}
{Af{\c{s}}ar}, M. \& {Ibano{\v{g}}lu}, C. 2008, \mnras, 391, 802

\bibitem[{{Althaus} {et~al.}(2013){Althaus}, {Miller Bertolami}, \&
  {C{\'o}rsico}}]{althaus}
{Althaus}, L.~G., {Miller Bertolami}, M.~M., \& {C{\'o}rsico}, A.~H. 2013,
  \aap, 557, A19

\bibitem[{{Astropy Collaboration} {et~al.}(2018){Astropy Collaboration},
  {Price-Whelan}, {Sip{\H{o}}cz}, {G{\"u}nther}, {Lim}, {Crawford}, {Conseil},
  {Shupe}, {Craig}, {Dencheva}, {Ginsburg}, {Vand erPlas}, {Bradley},
  {P{\'e}rez-Su{\'a}rez}, {de Val-Borro}, {Aldcroft}, {Cruz}, {Robitaille},
  {Tollerud}, {Ardelean}, {Babej}, {Bach}, {Bachetti}, {Bakanov}, {Bamford},
  {Barentsen}, {Barmby}, {Baumbach}, {Berry}, {Biscani}, {Boquien}, {Bostroem},
  {Bouma}, {Brammer}, {Bray}, {Breytenbach}, {Buddelmeijer}, {Burke},
  {Calderone}, {Cano Rodr{\'\i}guez}, {Cara}, {Cardoso}, {Cheedella}, {Copin},
  {Corrales}, {Crichton}, {D'Avella}, {Deil}, {Depagne}, {Dietrich}, {Donath},
  {Droettboom}, {Earl}, {Erben}, {Fabbro}, {Ferreira}, {Finethy}, {Fox},
  {Garrison}, {Gibbons}, {Goldstein}, {Gommers}, {Greco}, {Greenfield},
  {Groener}, {Grollier}, {Hagen}, {Hirst}, {Homeier}, {Horton}, {Hosseinzadeh},
  {Hu}, {Hunkeler}, {Ivezi{\'c}}, {Jain}, {Jenness}, {Kanarek}, {Kendrew},
  {Kern}, {Kerzendorf}, {Khvalko}, {King}, {Kirkby}, {Kulkarni}, {Kumar},
  {Lee}, {Lenz}, {Littlefair}, {Ma}, {Macleod}, {Mastropietro}, {McCully},
  {Montagnac}, {Morris}, {Mueller}, {Mumford}, {Muna}, {Murphy}, {Nelson},
  {Nguyen}, {Ninan}, {N{\"o}the}, {Ogaz}, {Oh}, {Parejko}, {Parley}, {Pascual},
  {Patil}, {Patil}, {Plunkett}, {Prochaska}, {Rastogi}, {Reddy Janga},
  {Sabater}, {Sakurikar}, {Seifert}, {Sherbert}, {Sherwood-Taylor}, {Shih},
  {Sick}, {Silbiger}, {Singanamalla}, {Singer}, {Sladen}, {Sooley},
  {Sornarajah}, {Streicher}, {Teuben}, {Thomas}, {Tremblay}, {Turner},
  {Terr{\'o}n}, {van Kerkwijk}, {de la Vega}, {Watkins}, {Weaver}, {Whitmore},
  {Woillez}, {Zabalza}, \& {Astropy Contributors}}]{astropy:2018}
{Astropy Collaboration}, {Price-Whelan}, A.~M., {Sip{\H{o}}cz}, B.~M., {et~al.}
  2018, \aj, 156, 123

\bibitem[{{Astropy Collaboration} {et~al.}(2013){Astropy Collaboration},
  {Robitaille}, {Tollerud}, {Greenfield}, {Droettboom}, {Bray}, {Aldcroft},
  {Davis}, {Ginsburg}, {Price-Whelan}, {Kerzendorf}, {Conley}, {Crighton},
  {Barbary}, {Muna}, {Ferguson}, {Grollier}, {Parikh}, {Nair}, {Unther},
  {Deil}, {Woillez}, {Conseil}, {Kramer}, {Turner}, {Singer}, {Fox}, {Weaver},
  {Zabalza}, {Edwards}, {Azalee Bostroem}, {Burke}, {Casey}, {Crawford},
  {Dencheva}, {Ely}, {Jenness}, {Labrie}, {Lim}, {Pierfederici}, {Pontzen},
  {Ptak}, {Refsdal}, {Servillat}, \& {Streicher}}]{astropy:2013}
{Astropy Collaboration}, {Robitaille}, T.~P., {Tollerud}, E.~J., {et~al.} 2013,
  \aap, 558, A33

\bibitem[{{Baran} {et~al.}(2021{\natexlab{a}}){Baran}, {{\O}stensen}, {Heber},
  {Irrgang}, {Sanjayan}, {Telting}, {Reed}, \& {Ostrowski}}]{baran21}
{Baran}, A.~S., {{\O}stensen}, R.~H., {Heber}, U., {et~al.} 2021{\natexlab{a}},
  \mnras, 503, 2157

\bibitem[{{Baran} {et~al.}(2021{\natexlab{b}}){Baran}, {Sahoo}, {Sanjayan}, \&
  {Ostrowski}}]{tess_north}
{Baran}, A.~S., {Sahoo}, S.~K., {Sanjayan}, S., \& {Ostrowski}, J.
  2021{\natexlab{b}}, \mnras

\bibitem[{{Barlow} {et~al.}(2022){Barlow}, {Corcoran}, {Parker}, {Kupfer},
  {N{\'e}meth}, {Hermes}, {Lopez}, {Frondorf}, {Vestal}, \&
  {Holden}}]{barlow22}
{Barlow}, B.~N., {Corcoran}, K.~A., {Parker}, I.~M., {et~al.} 2022, \apj, 928,
  20

\bibitem[{{Barlow} {et~al.}(2013){Barlow}, {Kilkenny}, {Drechsel}, {Dunlap},
  {O'Donoghue}, {Geier}, {O'Steen}, {Clemens}, {LaCluyze}, {Reichart},
  {Haislip}, {Nysewander}, \& {Ivarsen}}]{Barlow2013}
{Barlow}, B.~N., {Kilkenny}, D., {Drechsel}, H., {et~al.} 2013, \mnras, 430, 22

\bibitem[{{Bell} {et~al.}(2019){Bell}, {Kosakowski}, {Kilic}, {Green},
  {Latour}, {Baran}, {Charpinet}, {Handler}, {Pelisoli}, {Ratzloff}, \&
  {Silvotti}}]{bell19}
{Bell}, K.~J., {Kosakowski}, A., {Kilic}, M., {et~al.} 2019, Research Notes of
  the American Astronomical Society, 3, 81

\bibitem[{{Bloemen} {et~al.}(2011){Bloemen}, {Marsh}, {{\O}stensen},
  {Charpinet}, {Fontaine}, {Degroote}, {Heber}, {Kawaler}, {Aerts}, {Green},
  {Telting}, {Brassard}, {G{\"a}nsicke}, {Handler}, {Kurtz}, {Silvotti}, {Van
  Grootel}, {Lindberg}, {Pursimo}, {Wilson}, {Gilliland}, {Kjeldsen},
  {Christensen-Dalsgaard}, {Borucki}, {Koch}, {Jenkins}, \&
  {Klaus}}]{bloemen11}
{Bloemen}, S., {Marsh}, T.~R., {{\O}stensen}, R.~H., {et~al.} 2011, \mnras,
  410, 1787

\bibitem[{{Brown} {et~al.}(2008){Brown}, {Beers}, {Wilhelm}, {Allende Prieto},
  {Geller}, {Kenyon}, \& {Kurtz}}]{BROWN2008}
{Brown}, W.~R., {Beers}, T.~C., {Wilhelm}, R., {et~al.} 2008, \aj, 135, 564

\bibitem[{{Capitanio} {et~al.}(2017){Capitanio}, {Lallement}, {Vergely},
  {Elyajouri}, \& {Monreal-Ibero}}]{stilism2}
{Capitanio}, L., {Lallement}, R., {Vergely}, J.~L., {Elyajouri}, M., \&
  {Monreal-Ibero}, A. 2017, \aap, 606, A65

\bibitem[{{Chen} {et~al.}(1995){Chen}, {O'Donoghue}, {Stobie}, {Kilkenny},
  {Roberts}, \& {van Wyk}}]{Chen95}
{Chen}, A., {O'Donoghue}, D., {Stobie}, R.~S., {et~al.} 1995, \mnras, 275, 100

\bibitem[{{Copperwheat} {et~al.}(2011){Copperwheat}, {Morales-Rueda}, {Marsh},
  {Maxted}, \& {Heber}}]{Copperwheat11}
{Copperwheat}, C.~M., {Morales-Rueda}, L., {Marsh}, T.~R., {Maxted}, P.~F.~L.,
  \& {Heber}, U. 2011, \mnras, 415, 1381

\bibitem[{{Dorman} {et~al.}(1993){Dorman}, {Rood}, \&
  {O'Connell}}]{Dorman:1993}
{Dorman}, B., {Rood}, R.~T., \& {O'Connell}, R.~W. 1993, APJ, 419, 596

\bibitem[{{Drechsel} {et~al.}(2001){Drechsel}, {Heber}, {Napiwotzki},
  {{\O}stensen}, {Solheim}, {Johannessen}, {Schuh}, {Deetjen}, \&
  {Zola}}]{drechsel2001}
{Drechsel}, H., {Heber}, U., {Napiwotzki}, R., {et~al.} 2001, \aap, 379, 893

\bibitem[{{Drilling}(1985)}]{drilling85}
{Drilling}, J.~S. 1985, \apjl, 294, L107

\bibitem[{{Edelmann} {et~al.}(2005){Edelmann}, {Heber}, {Altmann}, {Karl}, \&
  {Lisker}}]{edelmann:2005}
{Edelmann}, H., {Heber}, U., {Altmann}, M., {Karl}, C., \& {Lisker}, T. 2005,
  \aap, 442, 1023

\bibitem[{{Edelmann} {et~al.}(2003){Edelmann}, {Heber}, {Hagen}, {Lemke},
  {Dreizler}, {Napiwotzki}, \& {Engels}}]{Edelmann2003}
{Edelmann}, H., {Heber}, U., {Hagen}, H.~J., {et~al.} 2003, \aap, 400, 939

\bibitem[{{For} {et~al.}(2010){For}, {Green}, {Fontaine}, {Drechsel}, {Shaw},
  {Dittmann}, {Fay}, {Francoeur}, {Laird}, {Moriyama}, {Morris},
  {Rodr{\'\i}guez-L{\'o}pez}, {Sierchio}, {Story}, {Strom}, {Wang}, {Adams},
  {Bolin}, {Eskew}, \& {Chayer}}]{For2010}
{For}, B.~Q., {Green}, E.~M., {Fontaine}, G., {et~al.} 2010, \apj, 708, 253

\bibitem[{{Gaia Collaboration} {et~al.}(2018){Gaia Collaboration}, {Brown},
  {Vallenari}, {Prusti}, {de Bruijne}, {Babusiaux}, {Bailer-Jones}, {Biermann},
  {Evans}, {Eyer}, {Jansen}, {Jordi}, {Klioner}, {Lammers}, {Lindegren},
  {Luri}, {Mignard}, {Panem}, {Pourbaix}, {Randich}, {Sartoretti}, {Siddiqui},
  {Soubiran}, {van Leeuwen}, {Walton}, {Arenou}, {Bastian}, {Cropper},
  {Drimmel}, {Katz}, {Lattanzi}, {Bakker}, {Cacciari}, {Casta{\~n}eda},
  {Chaoul}, {Cheek}, {De Angeli}, {Fabricius}, {Guerra}, {Holl}, {Masana},
  {Messineo}, {Mowlavi}, {Nienartowicz}, {Panuzzo}, {Portell}, {Riello},
  {Seabroke}, {Tanga}, {Th{\'e}venin}, {Gracia-Abril}, {Comoretto},
  {Garcia-Reinaldos}, {Teyssier}, {Altmann}, {Andrae}, {Audard},
  {Bellas-Velidis}, {Benson}, {Berthier}, {Blomme}, {Burgess}, {Busso},
  {Carry}, {Cellino}, {Clementini}, {Clotet}, {Creevey}, {Davidson}, {De
  Ridder}, {Delchambre}, {Dell'Oro}, {Ducourant},
  {Fern{\'a}ndez-Hern{\'a}ndez}, {Fouesneau}, {Fr{\'e}mat}, {Galluccio},
  {Garc{\'\i}a-Torres}, {Gonz{\'a}lez-N{\'u}{\~n}ez}, {Gonz{\'a}lez-Vidal},
  {Gosset}, {Guy}, {Halbwachs}, {Hambly}, {Harrison}, {Hern{\'a}ndez},
  {Hestroffer}, {Hodgkin}, {Hutton}, {Jasniewicz}, {Jean-Antoine-Piccolo},
  {Jordan}, {Korn}, {Krone-Martins}, {Lanzafame}, {Lebzelter}, {L{\"o}ffler},
  {Manteiga}, {Marrese}, {Mart{\'\i}n-Fleitas}, {Moitinho}, {Mora}, {Muinonen},
  {Osinde}, {Pancino}, {Pauwels}, {Petit}, {Recio-Blanco}, {Richards},
  {Rimoldini}, {Robin}, {Sarro}, {Siopis}, {Smith}, {Sozzetti}, {S{\"u}veges},
  {Torra}, {van Reeven}, {Abbas}, {Abreu Aramburu}, {Accart}, {Aerts},
  {Altavilla}, {{\'A}lvarez}, {Alvarez}, {Alves}, {Anderson}, {Andrei},
  {Anglada Varela}, {Antiche}, {Antoja}, {Arcay}, {Astraatmadja}, {Bach},
  {Baker}, {Balaguer-N{\'u}{\~n}ez}, {Balm}, {Barache}, {Barata}, {Barbato},
  {Barblan}, {Barklem}, {Barrado}, {Barros}, {Barstow}, {Bartholom{\'e}
  Mu{\~n}oz}, {Bassilana}, {Becciani}, {Bellazzini}, {Berihuete}, {Bertone},
  {Bianchi}, {Bienaym{\'e}}, {Blanco-Cuaresma}, {Boch}, {Boeche}, {Bombrun},
  {Borrachero}, {Bossini}, {Bouquillon}, {Bourda}, {Bragaglia}, {Bramante},
  {Breddels}, {Bressan}, {Brouillet}, {Br{\"u}semeister}, {Brugaletta},
  {Bucciarelli}, {Burlacu}, {Busonero}, {Butkevich}, {Buzzi}, {Caffau},
  {Cancelliere}, {Cannizzaro}, {Cantat-Gaudin}, {Carballo}, {Carlucci},
  {Carrasco}, {Casamiquela}, {Castellani}, {Castro-Ginard}, {Charlot},
  {Chemin}, {Chiavassa}, {Cocozza}, {Costigan}, {Cowell}, {Crifo}, {Crosta},
  {Crowley}, {Cuypers}, {Dafonte}, {Damerdji}, {Dapergolas}, {David}, {David},
  {de Laverny}, {De Luise}, {De March}, {de Martino}, {de Souza}, {de Torres},
  {Debosscher}, {del Pozo}, {Delbo}, {Delgado}, {Delgado}, {Di Matteo},
  {Diakite}, {Diener}, {Distefano}, {Dolding}, {Drazinos}, {Dur{\'a}n},
  {Edvardsson}, {Enke}, {Eriksson}, {Esquej}, {Eynard Bontemps}, {Fabre},
  {Fabrizio}, {Faigler}, {Falc{\~a}o}, {Farr{\`a}s Casas}, {Federici},
  {Fedorets}, {Fernique}, {Figueras}, {Filippi}, {Findeisen}, {Fonti},
  {Fraile}, {Fraser}, {Fr{\'e}zouls}, {Gai}, {Galleti}, {Garabato},
  {Garc{\'\i}a-Sedano}, {Garofalo}, {Garralda}, {Gavel}, {Gavras}, {Gerssen},
  {Geyer}, {Giacobbe}, {Gilmore}, {Girona}, {Giuffrida}, {Glass}, {Gomes},
  {Granvik}, {Gueguen}, {Guerrier}, {Guiraud}, {Guti{\'e}rrez-S{\'a}nchez},
  {Haigron}, {Hatzidimitriou}, {Hauser}, {Haywood}, {Heiter}, {Helmi}, {Heu},
  {Hilger}, {Hobbs}, {Hofmann}, {Holland}, {Huckle}, {Hypki}, {Icardi},
  {Jan{\ss}en}, {Jevardat de Fombelle}, {Jonker}, {Juh{\'a}sz}, {Julbe},
  {Karampelas}, {Kewley}, {Klar}, {Kochoska}, {Kohley}, {Kolenberg},
  {Kontizas}, {Kontizas}, {Koposov}, {Kordopatis}, {Kostrzewa-Rutkowska},
  {Koubsky}, {Lambert}, {Lanza}, {Lasne}, {Lavigne}, {Le Fustec}, {Le
  Poncin-Lafitte}, {Lebreton}, {Leccia}, {Leclerc}, {Lecoeur-Taibi},
  {Lenhardt}, {Leroux}, {Liao}, {Licata}, {Lindstr{\o}m}, {Lister}, {Livanou},
  {Lobel}, {L{\'o}pez}, {Managau}, {Mann}, {Mantelet}, {Marchal}, {Marchant},
  {Marconi}, {Marinoni}, {Marschalk{\'o}}, {Marshall}, {Martino}, {Marton},
  {Mary}, {Massari}, {Matijevi{\v{c}}}, {Mazeh}, {McMillan}, {Messina},
  {Michalik}, {Millar}, {Molina}, {Molinaro}, {Moln{\'a}r}, {Montegriffo},
  {Mor}, {Morbidelli}, {Morel}, {Morris}, {Mulone}, {Muraveva}, {Musella},
  {Nelemans}, {Nicastro}, {Noval}, {O'Mullane}, {Ord{\'e}novic},
  {Ord{\'o}{\~n}ez-Blanco}, {Osborne}, {Pagani}, {Pagano}, {Pailler},
  {Palacin}, {Palaversa}, {Panahi}, {Pawlak}, {Piersimoni}, {Pineau}, {Plachy},
  {Plum}, {Poggio}, {Poujoulet}, {Pr{\v{s}}a}, {Pulone}, {Racero}, {Ragaini},
  {Rambaux}, {Ramos-Lerate}, {Regibo}, {Reyl{\'e}}, {Riclet}, {Ripepi}, {Riva},
  {Rivard}, {Rixon}, {Roegiers}, {Roelens}, {Romero-G{\'o}mez}, {Rowell},
  {Royer}, {Ruiz-Dern}, {Sadowski}, {Sagrist{\`a} Sell{\'e}s}, {Sahlmann},
  {Salgado}, {Salguero}, {Sanna}, {Santana-Ros}, {Sarasso}, {Savietto},
  {Schultheis}, {Sciacca}, {Segol}, {Segovia}, {S{\'e}gransan}, {Shih},
  {Siltala}, {Silva}, {Smart}, {Smith}, {Solano}, {Solitro}, {Sordo}, {Soria
  Nieto}, {Souchay}, {Spagna}, {Spoto}, {Stampa}, {Steele},
  {Steidelm{\"u}ller}, {Stephenson}, {Stoev}, {Suess}, {Surdej}, {Szabados},
  {Szegedi-Elek}, {Tapiador}, {Taris}, {Tauran}, {Taylor}, {Teixeira},
  {Terrett}, {Teyssand ier}, {Thuillot}, {Titarenko}, {Torra Clotet}, {Turon},
  {Ulla}, {Utrilla}, {Uzzi}, {Vaillant}, {Valentini}, {Valette}, {van Elteren},
  {Van Hemelryck}, {van Leeuwen}, {Vaschetto}, {Vecchiato}, {Veljanoski},
  {Viala}, {Vicente}, {Vogt}, {von Essen}, {Voss}, {Votruba}, {Voutsinas},
  {Walmsley}, {Weiler}, {Wertz}, {Wevers}, {Wyrzykowski}, {Yoldas},
  {{\v{Z}}erjal}, {Ziaeepour}, {Zorec}, {Zschocke}, {Zucker}, {Zurbach}, \&
  {Zwitter}}]{gaia_dr2}
{Gaia Collaboration}, {Brown}, A.~G.~A., {Vallenari}, A., {et~al.} 2018, \aap,
  616, A1

\bibitem[{{Gaia Collaboration} {et~al.}(2021){Gaia Collaboration}, {Brown},
  {Vallenari}, {Prusti}, {de Bruijne}, {Babusiaux}, {Biermann}, {Creevey},
  {Evans}, {Eyer}, {Hutton}, {Jansen}, {Jordi}, {Klioner}, {Lammers},
  {Lindegren}, {Luri}, {Mignard}, {Panem}, {Pourbaix}, {Randich}, {Sartoretti},
  {Soubiran}, {Walton}, {Arenou}, {Bailer-Jones}, {Bastian}, {Cropper},
  {Drimmel}, {Katz}, {Lattanzi}, {van Leeuwen}, {Bakker}, {Cacciari},
  {Casta{\~n}eda}, {De Angeli}, {Ducourant}, {Fabricius}, {Fouesneau},
  {Fr{\'e}mat}, {Guerra}, {Guerrier}, {Guiraud}, {Jean-Antoine Piccolo},
  {Masana}, {Messineo}, {Mowlavi}, {Nicolas}, {Nienartowicz}, {Pailler},
  {Panuzzo}, {Riclet}, {Roux}, {Seabroke}, {Sordo}, {Tanga}, {Th{\'e}venin},
  {Gracia-Abril}, {Portell}, {Teyssier}, {Altmann}, {Andrae}, {Bellas-Velidis},
  {Benson}, {Berthier}, {Blomme}, {Brugaletta}, {Burgess}, {Busso}, {Carry},
  {Cellino}, {Cheek}, {Clementini}, {Damerdji}, {Davidson}, {Delchambre},
  {Dell'Oro}, {Fern{\'a}ndez-Hern{\'a}ndez}, {Galluccio}, {Garc{\'\i}a-Lario},
  {Garcia-Reinaldos}, {Gonz{\'a}lez-N{\'u}{\~n}ez}, {Gosset}, {Haigron},
  {Halbwachs}, {Hambly}, {Harrison}, {Hatzidimitriou}, {Heiter},
  {Hern{\'a}ndez}, {Hestroffer}, {Hodgkin}, {Holl}, {Jan{\ss}en}, {Jevardat de
  Fombelle}, {Jordan}, {Krone-Martins}, {Lanzafame}, {L{\"o}ffler}, {Lorca},
  {Manteiga}, {Marchal}, {Marrese}, {Moitinho}, {Mora}, {Muinonen}, {Osborne},
  {Pancino}, {Pauwels}, {Petit}, {Recio-Blanco}, {Richards}, {Riello},
  {Rimoldini}, {Robin}, {Roegiers}, {Rybizki}, {Sarro}, {Siopis}, {Smith},
  {Sozzetti}, {Ulla}, {Utrilla}, {van Leeuwen}, {van Reeven}, {Abbas}, {Abreu
  Aramburu}, {Accart}, {Aerts}, {Aguado}, {Ajaj}, {Altavilla}, {{\'A}lvarez},
  {{\'A}lvarez Cid-Fuentes}, {Alves}, {Anderson}, {Anglada Varela}, {Antoja},
  {Audard}, {Baines}, {Baker}, {Balaguer-N{\'u}{\~n}ez}, {Balbinot}, {Balog},
  {Barache}, {Barbato}, {Barros}, {Barstow}, {Bartolom{\'e}}, {Bassilana},
  {Bauchet}, {Baudesson-Stella}, {Becciani}, {Bellazzini}, {Bernet}, {Bertone},
  {Bianchi}, {Blanco-Cuaresma}, {Boch}, {Bombrun}, {Bossini}, {Bouquillon},
  {Bragaglia}, {Bramante}, {Breedt}, {Bressan}, {Brouillet}, {Bucciarelli},
  {Burlacu}, {Busonero}, {Butkevich}, {Buzzi}, {Caffau}, {Cancelliere},
  {C{\'a}novas}, {Cantat-Gaudin}, {Carballo}, {Carlucci}, {Carnerero},
  {Carrasco}, {Casamiquela}, {Castellani}, {Castro-Ginard}, {Castro Sampol},
  {Chaoul}, {Charlot}, {Chemin}, {Chiavassa}, {Cioni}, {Comoretto}, {Cooper},
  {Cornez}, {Cowell}, {Crifo}, {Crosta}, {Crowley}, {Dafonte}, {Dapergolas},
  {David}, {David}, {de Laverny}, {De Luise}, {De March}, {De Ridder}, {de
  Souza}, {de Teodoro}, {de Torres}, {del Peloso}, {del Pozo}, {Delbo},
  {Delgado}, {Delgado}, {Delisle}, {Di Matteo}, {Diakite}, {Diener},
  {Distefano}, {Dolding}, {Eappachen}, {Edvardsson}, {Enke}, {Esquej}, {Fabre},
  {Fabrizio}, {Faigler}, {Fedorets}, {Fernique}, {Fienga}, {Figueras},
  {Fouron}, {Fragkoudi}, {Fraile}, {Franke}, {Gai}, {Garabato},
  {Garcia-Gutierrez}, {Garc{\'\i}a-Torres}, {Garofalo}, {Gavras}, {Gerlach},
  {Geyer}, {Giacobbe}, {Gilmore}, {Girona}, {Giuffrida}, {Gomel}, {Gomez},
  {Gonzalez-Santamaria}, {Gonz{\'a}lez-Vidal}, {Granvik},
  {Guti{\'e}rrez-S{\'a}nchez}, {Guy}, {Hauser}, {Haywood}, {Helmi}, {Hidalgo},
  {Hilger}, {H{\l}adczuk}, {Hobbs}, {Holland}, {Huckle}, {Jasniewicz},
  {Jonker}, {Juaristi Campillo}, {Julbe}, {Karbevska}, {Kervella}, {Khanna},
  {Kochoska}, {Kontizas}, {Kordopatis}, {Korn}, {Kostrzewa-Rutkowska},
  {Kruszy{\'n}ska}, {Lambert}, {Lanza}, {Lasne}, {Le Campion}, {Le Fustec},
  {Lebreton}, {Lebzelter}, {Leccia}, {Leclerc}, {Lecoeur-Taibi}, {Liao},
  {Licata}, {Lindstr{\o}m}, {Lister}, {Livanou}, {Lobel}, {Madrero Pardo},
  {Managau}, {Mann}, {Marchant}, {Marconi}, {Marcos Santos}, {Marinoni},
  {Marocco}, {Marshall}, {Martin Polo}, {Mart{\'\i}n-Fleitas}, {Masip},
  {Massari}, {Mastrobuono-Battisti}, {Mazeh}, {McMillan}, {Messina},
  {Michalik}, {Millar}, {Mints}, {Molina}, {Molinaro}, {Moln{\'a}r},
  {Montegriffo}, {Mor}, {Morbidelli}, {Morel}, {Morris}, {Mulone}, {Munoz},
  {Muraveva}, {Murphy}, {Musella}, {Noval}, {Ord{\'e}novic}, {Orr{\`u}},
  {Osinde}, {Pagani}, {Pagano}, {Palaversa}, {Palicio}, {Panahi}, {Pawlak},
  {Pe{\~n}alosa Esteller}, {Penttil{\"a}}, {Piersimoni}, {Pineau}, {Plachy},
  {Plum}, {Poggio}, {Poretti}, {Poujoulet}, {Pr{\v{s}}a}, {Pulone}, {Racero},
  {Ragaini}, {Rainer}, {Raiteri}, {Rambaux}, {Ramos}, {Ramos-Lerate}, {Re
  Fiorentin}, {Regibo}, {Reyl{\'e}}, {Ripepi}, {Riva}, {Rixon}, {Robichon},
  {Robin}, {Roelens}, {Rohrbasser}, {Romero-G{\'o}mez}, {Rowell}, {Royer},
  {Rybicki}, {Sadowski}, {Sagrist{\`a} Sell{\'e}s}, {Sahlmann}, {Salgado},
  {Salguero}, {Samaras}, {Sanchez Gimenez}, {Sanna}, {Santove{\~n}a},
  {Sarasso}, {Schultheis}, {Sciacca}, {Segol}, {Segovia}, {S{\'e}gransan},
  {Semeux}, {Shahaf}, {Siddiqui}, {Siebert}, {Siltala}, {Slezak}, {Smart},
  {Solano}, {Solitro}, {Souami}, {Souchay}, {Spagna}, {Spoto}, {Steele},
  {Steidelm{\"u}ller}, {Stephenson}, {S{\"u}veges}, {Szabados}, {Szegedi-Elek},
  {Taris}, {Tauran}, {Taylor}, {Teixeira}, {Thuillot}, {Tonello}, {Torra},
  {Torra}, {Turon}, {Unger}, {Vaillant}, {van Dillen}, {Vanel}, {Vecchiato},
  {Viala}, {Vicente}, {Voutsinas}, {Weiler}, {Wevers}, {Wyrzykowski}, {Yoldas},
  {Yvard}, {Zhao}, {Zorec}, {Zucker}, {Zurbach}, \& {Zwitter}}]{gaia_edr3}
{Gaia Collaboration}, {Brown}, A.~G.~A., {Vallenari}, A., {et~al.} 2021, \aap,
  649, A1

\bibitem[{{Ge} {et~al.}(2022){Ge}, {Tout}, {Chen}, {Kruckow}, {Chen}, {Jiang},
  {Li}, {Liu}, \& {Han}}]{ge22}
{Ge}, H., {Tout}, C.~A., {Chen}, X., {et~al.} 2022, arXiv e-prints,
  arXiv:2205.14256

\bibitem[{{Geier}(2020)}]{geier:gaia}
{Geier}, S. 2020, \aap, 635, A193

\bibitem[{{Geier} {et~al.}(2010){Geier}, {Heber}, {Kupfer}, \&
  {Napiwotzki}}]{gd_687}
{Geier}, S., {Heber}, U., {Kupfer}, T., \& {Napiwotzki}, R. 2010, \aap, 515,
  A37

\bibitem[{{Geier} {et~al.}(2011{\natexlab{a}}){Geier}, {Hirsch}, {Tillich},
  {Maxted}, {Bentley}, {{\O}stensen}, {Heber}, {G{\"a}nsicke}, {Marsh},
  {Napiwotzki}, {Barlow}, \& {O'Toole}}]{geier:2011_2}
{Geier}, S., {Hirsch}, H., {Tillich}, A., {et~al.} 2011{\natexlab{a}}, \aap,
  530, A28

\bibitem[{{Geier} {et~al.}(2013){Geier}, {Marsh}, {Wang}, {Dunlap}, {Barlow},
  {Schaffenroth}, {Chen}, {Irrgang}, {Maxted}, {Ziegerer}, {Kupfer},
  {Miszalski}, {Heber}, {Han}, {Shporer}, {Telting}, {G{\"a}nsicke},
  {{\O}stensen}, {O'Toole}, \& {Napiwotzki}}]{cd-30}
{Geier}, S., {Marsh}, T.~R., {Wang}, B., {et~al.} 2013, \aap, 554, A54

\bibitem[{{Geier} {et~al.}(2011{\natexlab{b}}){Geier}, {Napiwotzki}, {Heber},
  \& {Nelemans}}]{egb5}
{Geier}, S., {Napiwotzki}, R., {Heber}, U., \& {Nelemans}, G.
  2011{\natexlab{b}}, \aap, 528, L16

\bibitem[{{Geier} {et~al.}(2014){Geier}, {{\O}stensen}, {Heber}, {Kupfer},
  {Maxted}, {Barlow}, {Vu{\v c}kovi{\'c}}, {Tillich}, {M{\"u}ller}, {Edelmann},
  {Classen}, \& {McLeod}}]{geier:2014}
{Geier}, S., {{\O}stensen}, R.~H., {Heber}, U., {et~al.} 2014, \aap, 562, A95

\bibitem[{{Geier} {et~al.}(2017){Geier}, {{\O}stensen}, {Nemeth}, {Gentile
  Fusillo}, {G{\"a}nsicke}, {Telting}, {Green}, \& {Schaffenroth}}]{Geier2017}
{Geier}, S., {{\O}stensen}, R.~H., {Nemeth}, P., {et~al.} 2017, \aap, 600, A50

\bibitem[{{Geier} {et~al.}(2019){Geier}, {Raddi}, {Gentile Fusillo}, \&
  {Marsh}}]{geier_gaia_catalog}
{Geier}, S., {Raddi}, R., {Gentile Fusillo}, N.~P., \& {Marsh}, T.~R. 2019,
  \aap, 621, A38

\bibitem[{{Gray}(2005)}]{gray}
{Gray}, D.~F. 2005, {The Observation and Analysis of Stellar Photospheres}

\bibitem[{{Han} {et~al.}(2003){Han}, {Podsiadlowski}, {Maxted}, \&
  {Marsh}}]{han:2003}
{Han}, Z., {Podsiadlowski}, P., {Maxted}, P.~F.~L., \& {Marsh}, T.~R. 2003,
  MNRAS, 341, 669

\bibitem[{{Han} {et~al.}(2002){Han}, {Podsiadlowski}, {Maxted}, {Marsh}, \&
  {Ivanova}}]{han:2002}
{Han}, Z., {Podsiadlowski}, P., {Maxted}, P.~F.~L., {Marsh}, T.~R., \&
  {Ivanova}, N. 2002, MNRAS, 336, 449

\bibitem[{{Heber}(2009)}]{heber:2009}
{Heber}, U. 2009, ARA{\rm \&}A, 47, 211

\bibitem[{{Heber}(2016)}]{heber:2016}
{Heber}, U. 2016, \pasp, 128, 082001

\bibitem[{{Heber} {et~al.}(2004){Heber}, {Drechsel}, {{\O}stensen}, {Karl},
  {Napiwotzki}, {Altmann}, {Cordes}, {Solheim}, {Voss}, {Koester}, \&
  {Folkes}}]{heber2004}
{Heber}, U., {Drechsel}, H., {{\O}stensen}, R., {et~al.} 2004, \aap, 420, 251

\bibitem[{{Heber} {et~al.}(2018){Heber}, {Irrgang}, \&
  {Schaffenroth}}]{heber:2018}
{Heber}, U., {Irrgang}, A., \& {Schaffenroth}, J. 2018, Open Astronomy, 27, 35

\bibitem[{{Hillwig} {et~al.}(2017){Hillwig}, {Frew}, {Reindl}, {Rotter},
  {Webb}, \& {Margheim}}]{todd}
{Hillwig}, T.~C., {Frew}, D.~J., {Reindl}, N., {et~al.} 2017, \aj, 153, 24

\bibitem[{{Howell} {et~al.}(2014){Howell}, {Sobeck}, {Haas}, {Still},
  {Barclay}, {Mullally}, {Troeltzsch}, {Aigrain}, {Bryson}, {Caldwell},
  {Chaplin}, {Cochran}, {Huber}, {Marcy}, {Miglio}, {Najita}, {Smith},
  {Twicken}, \& {Fortney}}]{k2}
{Howell}, S.~B., {Sobeck}, C., {Haas}, M., {et~al.} 2014, \pasp, 126, 398

\bibitem[{{Irrgang} {et~al.}(2021){Irrgang}, {Geier}, {Heber}, {Kupfer},
  {El-Badry}, \& {Bloemen}}]{sed_andreas}
{Irrgang}, A., {Geier}, S., {Heber}, U., {et~al.} 2021, \aap, 650, A102

\bibitem[{{Ivanova} {et~al.}(2013){Ivanova}, {Justham}, {Chen}, {De Marco},
  {Fryer}, {Gaburov}, {Ge}, {Glebbeek}, {Han}, {Li}, {Lu}, {Marsh},
  {Podsiadlowski}, {Potter}, {Soker}, {Taam}, {Tauris}, {van den Heuvel}, \&
  {Webbink}}]{ivanova}
{Ivanova}, N., {Justham}, S., {Chen}, X., {et~al.} 2013, \aapr, 21, 59

\bibitem[{{Jeffery} \& {Ramsay}(2014)}]{jeffrey:2014}
{Jeffery}, C.~S. \& {Ramsay}, G. 2014, \mnras, 442, L61

\bibitem[{{Kawka} {et~al.}(2010){Kawka}, {Vennes}, {N{\'e}meth}, {Kraus}, \&
  {Kub{\'a}t}}]{kawka:2010}
{Kawka}, A., {Vennes}, S., {N{\'e}meth}, P., {Kraus}, M., \& {Kub{\'a}t}, J.
  2010, \mnras, 408, 992

\bibitem[{{Kawka} {et~al.}(2015){Kawka}, {Vennes}, {O'Toole}, {N{\'e}meth},
  {Burton}, {Kotze}, \& {Buckley}}]{kawka:2015}
{Kawka}, A., {Vennes}, S., {O'Toole}, S., {et~al.} 2015, \mnras, 450, 3514

\bibitem[{{Kilkenny} {et~al.}(2010){Kilkenny}, {Koen}, \&
  {Worters}}]{Kilkenny2010}
{Kilkenny}, D., {Koen}, C., \& {Worters}, H. 2010, \mnras, 404, 376

\bibitem[{{Kilkenny} {et~al.}(2015){Kilkenny}, {O'Donoghue}, {Worters}, {Koen},
  {Hambly}, \& {MacGillivray}}]{Kilkenny2015}
{Kilkenny}, D., {O'Donoghue}, D., {Worters}, H.~L., {et~al.} 2015, \mnras, 453,
  1879

\bibitem[{{Kilkenny} \& {Stone}(1988)}]{Kilkenny1988}
{Kilkenny}, D. \& {Stone}, L.~E. 1988, \mnras, 234, 1011

\bibitem[{{Kilkenny} {et~al.}(2016){Kilkenny}, {Worters}, {O'Donoghue}, {Koen},
  {Koen}, {Hambly}, {MacGillivray}, \& {Stobie}}]{Kilkenny2016}
{Kilkenny}, D., {Worters}, H.~L., {O'Donoghue}, D., {et~al.} 2016, \mnras, 459,
  4343

\bibitem[{{Koen}(2009)}]{koen_2009}
{Koen}, C. 2009, \mnras, 395, 979

\bibitem[{{Koen} {et~al.}(1999){Koen}, {O'Donoghue}, {Kilkenny}, {Stobie}, \&
  {Saffer}}]{Koen1999}
{Koen}, C., {O'Donoghue}, D., {Kilkenny}, D., {Stobie}, R.~S., \& {Saffer},
  R.~A. 1999, \mnras, 306, 213

\bibitem[{{Krzesinski} \& {Balona}(2022)}]{krzesinski}
{Krzesinski}, J. \& {Balona}, L.~A. 2022, arXiv e-prints, arXiv:2204.01604

\bibitem[{{Kupfer} {et~al.}(2019){Kupfer}, {Bauer}, {Burdge}, {Bellm},
  {Bildsten}, {Fuller}, {Hermes}, {Kulkarni}, {Prince}, {van Roestel},
  {Dekany}, {Duev}, {Feeney}, {Giomi}, {Graham}, {Kaye}, {Laher}, {Masci},
  {Porter}, {Riddle}, {Shupe}, {Smith}, {Soumagnac}, {Szkody}, \&
  {Ward}}]{kupfer2019}
{Kupfer}, T., {Bauer}, E.~B., {Burdge}, K.~B., {et~al.} 2019, \apjl, 878, L35

\bibitem[{{Kupfer} {et~al.}(2020{\natexlab{a}}){Kupfer}, {Bauer}, {Burdge},
  {Roestel}, {Bellm}, {Fuller}, {Hermes}, {Marsh}, {Bildsten}, {Kulkarni},
  {Phinney}, {Prince}, {Szkody}, {Yao}, {Irrgang}, {Heber}, {Schneider},
  {Dhillon}, {Murawski}, {Drake}, {Duev}, {Feeney}, {Graham}, {Laher},
  {Littlefair}, {Mahabal}, {Masci}, {Porter}, {Reiley}, {Rodriguez},
  {Rusholme}, {Shupe}, \& {Soumagnac}}]{kupfer:20a}
{Kupfer}, T., {Bauer}, E.~B., {Burdge}, K.~B., {et~al.} 2020{\natexlab{a}},
  \apjl, 898, L25

\bibitem[{{Kupfer} {et~al.}(2020{\natexlab{b}}){Kupfer}, {Bauer}, {Marsh},
  {Roestel}, {Bellm}, {Burdge}, {Coughlin}, {Fuller}, {Hermes}, {Bildsten},
  {Kulkarni}, {Prince}, {Szkody}, {Dhillon}, {Murawski}, {Burruss}, {Dekany},
  {Delacroix}, {Drake}, {Duev}, {Feeney}, {Graham}, {Kaplan}, {Laher},
  {Littlefair}, {Masci}, {Riddle}, {Rusholme}, {Serabyn}, {Smith}, {Shupe}, \&
  {Soumagnac}}]{kupfer20}
{Kupfer}, T., {Bauer}, E.~B., {Marsh}, T.~R., {et~al.} 2020{\natexlab{b}},
  \apj, 891, 45

\bibitem[{{Kupfer} {et~al.}(2022){Kupfer}, {Bauer}, {van Roestel}, {Bellm},
  {Bildsten}, {Fuller}, {Prince}, {Heber}, {Geier}, {Green}, {Kulkarni},
  {Bloemen}, {Laher}, {Rusholme}, \& {Schneider}}]{kupfer:22}
{Kupfer}, T., {Bauer}, E.~B., {van Roestel}, J., {et~al.} 2022, \apjl, 925, L12

\bibitem[{{Kupfer} {et~al.}(2015){Kupfer}, {Geier}, {Heber}, {{\O}stensen},
  {Barlow}, {Maxted}, {Heuser}, {Schaffenroth}, \& {G{\"a}nsicke}}]{Kupfer2015}
{Kupfer}, T., {Geier}, S., {Heber}, U., {et~al.} 2015, \aap, 576, A44

\bibitem[{{Kupfer} {et~al.}(2017{\natexlab{a}}){Kupfer}, {Ramsay}, {van
  Roestel}, {Brooks}, {MacFarlane}, {Toma}, {Groot}, {Woudt}, {Bildsten},
  {Marsh}, {Green}, {Breedt}, {Kilkenny}, {Freudenthal}, {Geier}, {Heber},
  {Bagnulo}, {Blagorodnova}, {Buckley}, {Dhillon}, {Kulkarni}, {Lunnan}, \&
  {Prince}}]{kupfer17b}
{Kupfer}, T., {Ramsay}, G., {van Roestel}, J., {et~al.} 2017{\natexlab{a}},
  \apj, 851, 28

\bibitem[{{Kupfer} {et~al.}(2017{\natexlab{b}}){Kupfer}, {van Roestel},
  {Brooks}, {Geier}, {Marsh}, {Groot}, {Bloemen}, {Prince}, {Bellm}, {Heber},
  {Bildsten}, {Miller}, {Dyer}, {Dhillon}, {Green}, {Irawati}, {Laher},
  {Littlefair}, {Shupe}, {Steidel}, {Rattansoon}, \& {Pettini}}]{kupfer:2017}
{Kupfer}, T., {van Roestel}, J., {Brooks}, J., {et~al.} 2017{\natexlab{b}},
  \apj, 835, 131

\bibitem[{{Lallement} {et~al.}(2014){Lallement}, {Vergely}, {Valette},
  {Puspitarini}, {Eyer}, \& {Casagrande}}]{stilism}
{Lallement}, R., {Vergely}, J.-L., {Valette}, B., {et~al.} 2014, \aap, 561, A91

\bibitem[{{Lamontagne} {et~al.}(2000){Lamontagne}, {Demers}, {Wesemael},
  {Fontaine}, \& {Irwin}}]{Lamontagne2000}
{Lamontagne}, R., {Demers}, S., {Wesemael}, F., {Fontaine}, G., \& {Irwin},
  M.~J. 2000, \aj, 119, 241

\bibitem[{{Latour} {et~al.}(2014){Latour}, {Fontaine}, \& {Green}}]{latour14}
{Latour}, M., {Fontaine}, G., \& {Green}, E. 2014, in Astronomical Society of
  the Pacific Conference Series, Vol. 481, 6th Meeting on Hot Subdwarf Stars
  and Related Objects, ed. V.~{van Grootel}, E.~{Green}, G.~{Fontaine}, \&
  S.~{Charpinet}, 91

\bibitem[{{Lei} {et~al.}(2018){Lei}, {Zhao}, {N{\'e}meth}, \& {Zhao}}]{Lei2018}
{Lei}, Z., {Zhao}, J., {N{\'e}meth}, P., \& {Zhao}, G. 2018, \apj, 868, 70

\bibitem[{{Lightkurve Collaboration} {et~al.}(2018){Lightkurve Collaboration},
  {Cardoso}, {Hedges}, {Gully-Santiago}, {Saunders}, {Cody}, {Barclay}, {Hall},
  {Sagear}, {Turtelboom}, {Zhang}, {Tzanidakis}, {Mighell}, {Coughlin}, {Bell},
  {Berta-Thompson}, {Williams}, {Dotson}, \& {Barentsen}}]{lightkurve}
{Lightkurve Collaboration}, {Cardoso}, J.~V.~d.~M., {Hedges}, C., {et~al.}
  2018, {Lightkurve: Kepler and TESS time series analysis in Python},
  Astrophysics Source Code Library

\bibitem[{{Lomb}(1976)}]{lomb}
{Lomb}, N.~R. 1976, \apss, 39, 447

\bibitem[{{Lynas-Gray}(2021)}]{lynasgray21}
{Lynas-Gray}, A.~E. 2021, Frontiers in Astronomy and Space Sciences, 8, 19

\bibitem[{{Maxted} {et~al.}(2001){Maxted}, {Heber}, {Marsh}, \&
  {North}}]{Maxted2001}
{Maxted}, P.~F.~L., {Heber}, U., {Marsh}, T.~R., \& {North}, R.~C. 2001,
  \mnras, 326, 1391

\bibitem[{{Menzies} \& {Marang}(1986)}]{menzies:1986}
{Menzies}, J.~W. \& {Marang}, F. 1986, in IAU Symposium, Vol. 118,
  Instrumentation and Research Programmes for Small Telescopes, ed. J.~B.
  {Hearnshaw} \& P.~L. {Cottrell}, 305

\bibitem[{{Mickaelian}(2008)}]{mickelian2008}
{Mickaelian}, A.~M. 2008, \aj, 136, 946

\bibitem[{{M\"oller}(2021)}]{moeller}
{M\"oller}, L. 2021

\bibitem[{{Morales-Rueda} {et~al.}(2005){Morales-Rueda}, {Maxted}, {Marsh},
  {Kilkenny}, \& {O'Donoghue}}]{morales-rueda05}
{Morales-Rueda}, L., {Maxted}, P.~F.~L., {Marsh}, T.~R., {Kilkenny}, D., \&
  {O'Donoghue}, D. 2005, in Astronomical Society of the Pacific Conference
  Series, Vol. 334, 14th European Workshop on White Dwarfs, ed. D.~{Koester} \&
  S.~{Moehler}, 333

\bibitem[{{N{\'e}meth} {et~al.}(2012){N{\'e}meth}, {Kawka}, \&
  {Vennes}}]{nemeth:2012}
{N{\'e}meth}, P., {Kawka}, A., \& {Vennes}, S. 2012, \mnras, 427, 2180

\bibitem[{{O'Donoghue} {et~al.}(2013){O'Donoghue}, {Kilkenny}, {Koen},
  {Hambly}, {MacGillivray}, \& {Stobie}}]{Donogue2013}
{O'Donoghue}, D., {Kilkenny}, D., {Koen}, C., {et~al.} 2013, \mnras, 431, 240

\bibitem[{{Orosz} \& {Wade}(1999)}]{orosz99}
{Orosz}, J.~A. \& {Wade}, R.~A. 1999, \mnras, 310, 773

\bibitem[{{{\O}stensen} {et~al.}(2013){{\O}stensen}, {Geier}, {Schaffenroth},
  {Telting}, {Bloemen}, {N{\'e}meth}, {Beck}, {Lombaert}, {P{\'a}pics},
  {Tillich}, {Ziegerer}, {Fox Machado}, {Littlefair}, {Dhillon}, {Aerts},
  {Heber}, {Maxted}, {G{\"a}nsicke}, \& {Marsh}}]{oestensen2013}
{{\O}stensen}, R.~H., {Geier}, S., {Schaffenroth}, V., {et~al.} 2013, \aap,
  559, A35

\bibitem[{{{\O}stensen} {et~al.}(2010{\natexlab{a}}){{\O}stensen}, {Green},
  {Bloemen}, {Marsh}, {Laird}, {Morris}, {Moriyama}, {Oreiro}, {Reed},
  {Kawaler}, {Aerts}, {Vu{\v{c}}kovi{\'c}}, {Degroote}, {Telting}, {Kjeldsen},
  {Gilliland}, {Christensen-Dalsgaard}, {Borucki}, \& {Koch}}]{oestensen10}
{{\O}stensen}, R.~H., {Green}, E.~M., {Bloemen}, S., {et~al.}
  2010{\natexlab{a}}, \mnras, 408, L51

\bibitem[{{{\O}stensen} {et~al.}(2010{\natexlab{b}}){{\O}stensen}, {Green},
  {Bloemen}, {Marsh}, {Laird}, {Morris}, {Moriyama}, {Oreiro}, {Reed},
  {Kawaler}, {Aerts}, {Vu{\v{c}}kovi{\'c}}, {Degroote}, {Telting}, {Kjeldsen},
  {Gilliland}, {Christensen-Dalsgaard}, {Borucki}, \& {Koch}}]{oestensen_2010}
{{\O}stensen}, R.~H., {Green}, E.~M., {Bloemen}, S., {et~al.}
  2010{\natexlab{b}}, \mnras, 408, L51

\bibitem[{{{\O}stensen} {et~al.}(2010{\natexlab{c}}){{\O}stensen}, {Oreiro},
  {Solheim}, {Heber}, {Silvotti}, {Gonz{\'a}lez-P{\'e}rez}, {Ulla}, {P{\'e}rez
  Hern{\'a}ndez}, {Rodr{\'\i}guez-L{\'o}pez}, \& {Telting}}]{Oestensen2010}
{{\O}stensen}, R.~H., {Oreiro}, R., {Solheim}, J.~E., {et~al.}
  2010{\natexlab{c}}, \aap, 513, A6

\bibitem[{{Pawar}(2020)}]{tilak}
{Pawar}, T. 2020

\bibitem[{{Pelisoli} {et~al.}(2021){Pelisoli}, {Neunteufel}, {Geier}, {Kupfer},
  {Heber}, {Irrgang}, {Schneider}, {Bastian}, {van Roestel}, {Schaffenroth}, \&
  {Barlow}}]{pelisoli21}
{Pelisoli}, I., {Neunteufel}, P., {Geier}, S., {et~al.} 2021, Nature Astronomy,
  5, 1052

\bibitem[{{Pelisoli} {et~al.}(2020){Pelisoli}, {Vos}, {Geier}, {Schaffenroth},
  \& {Baran}}]{pelisoli:2020}
{Pelisoli}, I., {Vos}, J., {Geier}, S., {Schaffenroth}, V., \& {Baran}, A.~S.
  2020, \aap, 642, A180

\bibitem[{{Penoyre} {et~al.}(2022){Penoyre}, {Belokurov}, \&
  {Evans}}]{penoyre22}
{Penoyre}, Z., {Belokurov}, V., \& {Evans}, N.~W. 2022, \mnras, 513, 2437

\bibitem[{{Pribulla} {et~al.}(2013){Pribulla}, {Dimitrov}, {Kjurkchieva},
  {Kohl}, {Kundra}, {Ohlert}, {Perdelwitz}, {Srdoc}, \& {Vanko}}]{Pribulla2013}
{Pribulla}, T., {Dimitrov}, D., {Kjurkchieva}, D., {et~al.} 2013, Information
  Bulletin on Variable Stars, 6067, 1

\bibitem[{{Ratzloff} {et~al.}(2019){Ratzloff}, {Barlow}, {Kupfer}, {Corcoran},
  {Geier}, {Bauer}, {Corbett}, {Howard}, {Glazier}, \& {Law}}]{evr01}
{Ratzloff}, J.~K., {Barlow}, B.~N., {Kupfer}, T., {et~al.} 2019, \apj, 883, 51

\bibitem[{{Ratzloff} {et~al.}(2020{\natexlab{a}}){Ratzloff}, {Barlow},
  {N{\'e}meth}, {Corbett}, {Walser}, {Galliher}, {Glazier}, {Howard}, \&
  {Law}}]{Evr03}
{Ratzloff}, J.~K., {Barlow}, B.~N., {N{\'e}meth}, P., {et~al.}
  2020{\natexlab{a}}, \apj, 890, 126

\bibitem[{{Ratzloff} {et~al.}(2020{\natexlab{b}}){Ratzloff}, {Kupfer},
  {Barlow}, {Schneider}, {Marsh}, {Heber}, {Corcoran}, {Bauer},
  {H{\"a}mmerich}, {Corbett}, {Glazier}, {Howard}, \& {Law}}]{evr04}
{Ratzloff}, J.~K., {Kupfer}, T., {Barlow}, B.~N., {et~al.} 2020{\natexlab{b}},
  \apj, 902, 92

\bibitem[{{Reed} {et~al.}(2004){Reed}, {Kawaler}, {Zola}, {Jiang}, {Dreizler},
  {Schuh}, {Deetjen}, {Kalytis}, {Mei{\v{s}}tas}, {Janulis},
  {Ali{\v{s}}auskas}, {Krzesi{\'n}ski}, {Vuckovic}, {Moskalik}, {Og{\l}oza},
  {Baran}, {Stachowski}, {Kurtz}, {Gonz{\'a}lez P{\'e}rez}, {Mukadam},
  {Watson}, {Koen}, {Bradley}, {Cunha}, {Kilic}, {Klumpe}, {Carlton},
  {Handler}, {Kilkenny}, {Riddle}, {Dolez}, {Vauclair}, {Chevreton}, {Wood},
  {Grauer}, {Bromage}, {Solheim}, {{\O}stensen}, {Ulla}, {Burleigh}, {Good},
  {H{\"u}rkal}, {Anderson}, \& {Pakstiene}}]{Reed2004}
{Reed}, M.~D., {Kawaler}, S.~D., {Zola}, S., {et~al.} 2004, \mnras, 348, 1164

\bibitem[{{Reed} {et~al.}(2020){Reed}, {Yeager}, {Vos}, {Telting},
  {{\O}stensen}, {Slayton}, {Baran}, \& {Jeffery}}]{reed20}
{Reed}, M.~D., {Yeager}, M., {Vos}, J., {et~al.} 2020, \mnras, 492, 5202

\bibitem[{{Ricker} {et~al.}(2015){Ricker}, {Winn}, {Vanderspek}, {Latham},
  {Bakos}, {Bean}, {Berta-Thompson}, {Brown}, {Buchhave}, {Butler}, {Butler},
  {Chaplin}, {Charbonneau}, {Christensen-Dalsgaard}, {Clampin}, {Deming},
  {Doty}, {De Lee}, {Dressing}, {Dunham}, {Endl}, {Fressin}, {Ge}, {Henning},
  {Holman}, {Howard}, {Ida}, {Jenkins}, {Jernigan}, {Johnson}, {Kaltenegger},
  {Kawai}, {Kjeldsen}, {Laughlin}, {Levine}, {Lin}, {Lissauer}, {MacQueen},
  {Marcy}, {McCullough}, {Morton}, {Narita}, {Paegert}, {Palle}, {Pepe},
  {Pepper}, {Quirrenbach}, {Rinehart}, {Sasselov}, {Sato}, {Seager},
  {Sozzetti}, {Stassun}, {Sullivan}, {Szentgyorgyi}, {Torres}, {Udry}, \&
  {Villasenor}}]{TESS}
{Ricker}, G.~R., {Winn}, J.~N., {Vanderspek}, R., {et~al.} 2015, Journal of
  Astronomical Telescopes, Instruments, and Systems, 1, 014003

\bibitem[{{Rodr{\'\i}guez-L{\'o}pez} {et~al.}(2007){Rodr{\'\i}guez-L{\'o}pez},
  {Ulla}, \& {Garrido}}]{Rodriguez-Lopez2007}
{Rodr{\'\i}guez-L{\'o}pez}, C., {Ulla}, A., \& {Garrido}, R. 2007, \mnras, 379,
  1123

\bibitem[{{Sahoo} {et~al.}(2020){Sahoo}, {Baran}, {Sanjayan}, \&
  {Ostrowski}}]{tess_south}
{Sahoo}, S.~K., {Baran}, A.~S., {Sanjayan}, S., \& {Ostrowski}, J. 2020,
  \mnras, 499, 5508

\bibitem[{{Scargle}(1982)}]{scargle}
{Scargle}, J.~D. 1982, \apj, 263, 835

\bibitem[{{Schaffenroth} {et~al.}(2019){Schaffenroth}, {Barlow}, {Geier},
  {Vu{\v{c}}kovi{\'c}}, {Kilkenny}, {Wolz}, {Kupfer}, {Heber}, {Drechsel},
  {Kimeswenger}, {Marsh}, {Wolf}, {Pelisoli}, {Freudenthal}, {Dreizler},
  {Kreuzer}, \& {Ziegerer}}]{erebos}
{Schaffenroth}, V., {Barlow}, B.~N., {Geier}, S., {et~al.} 2019, \aap, 630, A80

\bibitem[{{Schaffenroth} {et~al.}(2021){Schaffenroth}, {Casewell}, {Schneider},
  {Kilkenny}, {Geier}, {Heber}, {Irrgang}, {Przybilla}, {Marsh}, {Littlefair},
  \& {Dhillon}}]{Schaffenroth2020}
{Schaffenroth}, V., {Casewell}, S.~L., {Schneider}, D., {et~al.} 2021, \mnras,
  501, 3847

\bibitem[{{Schaffenroth} {et~al.}(2014{\natexlab{a}}){Schaffenroth}, {Classen},
  {Nagel}, {Geier}, {Koen}, {Heber}, \& {Edelmann}}]{Schaffenroth2014_II}
{Schaffenroth}, V., {Classen}, L., {Nagel}, K., {et~al.} 2014{\natexlab{a}},
  \aap, 570, A70

\bibitem[{{Schaffenroth} {et~al.}(2013){Schaffenroth}, {Geier}, {Drechsel},
  {Heber}, {Wils}, {{\O}stensen}, {Maxted}, \& {di Scala}}]{Schaffenroth2013}
{Schaffenroth}, V., {Geier}, S., {Drechsel}, H., {et~al.} 2013, \aap, 553, A18

\bibitem[{{Schaffenroth} {et~al.}(2018){Schaffenroth}, {Geier}, {Heber},
  {Gerber}, {Schneider}, {Ziegerer}, \& {Cordes}}]{muchfuss_photo}
{Schaffenroth}, V., {Geier}, S., {Heber}, U., {et~al.} 2018, \aap, 614, A77

\bibitem[{{Schaffenroth} {et~al.}(2014{\natexlab{b}}){Schaffenroth}, {Geier},
  {Heber}, {Kupfer}, {Ziegerer}, {Heuser}, {Classen}, \&
  {Cordes}}]{Schaffenroth2014_I}
{Schaffenroth}, V., {Geier}, S., {Heber}, U., {et~al.} 2014{\natexlab{b}},
  \aap, 564, A98

\bibitem[{{Schindewolf} {et~al.}(2015){Schindewolf}, {Levitan}, {Heber},
  {Drechsel}, {Schaffenroth}, {Kupfer}, \& {Prince}}]{schindewolf}
{Schindewolf}, M., {Levitan}, D., {Heber}, U., {et~al.} 2015, \aap, 580, A117

\bibitem[{{Silvotti} {et~al.}(2012){Silvotti}, {{\O}stensen}, {Bloemen},
  {Telting}, {Heber}, {Oreiro}, {Reed}, {Farris}, {O'Toole}, {Lanteri},
  {Degroote}, {Hu}, {Baran}, {Hermes}, {Althaus}, {Marsh}, {Charpinet}, {Li},
  {Morris}, \& {Sanderfer}}]{silvotti12}
{Silvotti}, R., {{\O}stensen}, R.~H., {Bloemen}, S., {et~al.} 2012, \mnras,
  424, 1752

\bibitem[{{Vos} {et~al.}(2018){Vos}, {N{\'e}meth}, {Vu{\v{c}}kovi{\'c}},
  {{\O}stensen}, \& {Parsons}}]{vos:2018}
{Vos}, J., {N{\'e}meth}, P., {Vu{\v{c}}kovi{\'c}}, M., {{\O}stensen}, R., \&
  {Parsons}, S. 2018, \mnras, 473, 693

\bibitem[{{Vu{\v c}kovi{\'c}} {et~al.}(2007){Vu{\v c}kovi{\'c}}, {Aerts},
  {{\"O}stensen}, {Nelemans}, {Hu}, {Jeffery}, {Dhillon}, \& {Marsh}}]{nyvir}
{Vu{\v c}kovi{\'c}}, M., {Aerts}, C., {{\"O}stensen}, R., {et~al.} 2007, \aap,
  471, 605

\bibitem[{{Vu{\v{c}}kovi{\'c}} {et~al.}(2014){Vu{\v{c}}kovi{\'c}}, {Bloemen},
  \& {{\"O}stensen}}]{vuckovic14}
{Vu{\v{c}}kovi{\'c}}, M., {Bloemen}, S., \& {{\"O}stensen}, R. 2014, in
  Astronomical Society of the Pacific Conference Series, Vol. 481, 6th Meeting
  on Hot Subdwarf Stars and Related Objects, ed. V.~{van Grootel}, E.~{Green},
  G.~{Fontaine}, \& S.~{Charpinet}, 259

\bibitem[{{Vu{\v{c}}kovi{\'c}} {et~al.}(2009){Vu{\v{c}}kovi{\'c}},
  {{\O}stensen}, {Aerts}, {Telting}, {Heber}, \& {Oreiro}}]{Vuckovic2009}
{Vu{\v{c}}kovi{\'c}}, M., {{\O}stensen}, R.~H., {Aerts}, C., {et~al.} 2009,
  \aap, 505, 239

\bibitem[{{Vu{\v{c}}kovi{\'c}} {et~al.}(2016){Vu{\v{c}}kovi{\'c}},
  {{\O}stensen}, {N{\'e}meth}, {Bloemen}, \& {P{\'a}pics}}]{Vuckovic2016}
{Vu{\v{c}}kovi{\'c}}, M., {{\O}stensen}, R.~H., {N{\'e}meth}, P., {Bloemen},
  S., \& {P{\'a}pics}, P.~I. 2016, \aap, 586, A146

\end{thebibliography}
\bibliographystyle{aa}

\begin{appendix}

%\newpage\onecolumn

%\newpage\twocolumn

\newpage\onecolumn

%\section{Atmospheric and absolute parameters of the sdB primaries}
\section{Parameters of the close sdB binaries in TESS}
\begin{longtable}{lllllll}
\caption{Atmospheric and absolute parameters of the sdB binaries with spectroscopic parameters and with space-based light curves determined by spectroscopy and spectral energy distribution fitting together with the \textit{Gaia} parallax.}
\label{sed_known}\\
\hline
%\begin{tabular}{llllllll}
target & $T_{\rm eff,spec}$ & $\log{g}_{\rm spec}$ & $T_{\rm eff,sed}$  & $M_{\rm sed}$ & $L_{\rm sed}$ & $R_{\rm sed}$ \\
& [K ] & [cgs] &  [K]  & [M$_\odot$] & [L$_\odot$] & [R$_\odot$] \\\hline
\multicolumn{7}{c}{Reflection effect systems}\\\hline
KPD2215+5037 & 29600 $\pm$ 1000& 5.64$\pm$ 0.10 & $27000^{+7000}_{-6000}$ & $0.445^{+0.063}_{-0.055}$ & $19.9^{+1.9}_{-1.7}$ & $0.170^{+0.005}_{-0.005}$ \\
JL82 & 26500$\pm$ 500 & 5.22$\pm$ 0.1  & $26000^{+500}_{-600}$ & $0.294^{+0.04}_{-0.034}$ & $21^{+1.9}_{-2.0}$ & $0.222^{+0.006}_{-0.006}$ \\
HE0230-4323 & 31552$\pm$ 500 & 5.60$\pm$ 0.07& $34000^{+8000}_{-5000}$ & $0.635^{+0.118}_{-0.094}$ & $40^{+6}_{-5}$ & $0.211^{+0.014}_{-0.012}$ \\
V1405Ori & 35100$\pm$ 800 & 5.66$\pm$ 0.11 & $27000^{+2300}_{-2700}$ & $0.44^{+0.06}_{-0.055}$ & $36^{+5}_{-5}$ & $0.166^{+0.006}_{-0.006}$ \\
\end{longtable}

\textit{Notes:} spectroscopic parameters can be found in \citet[][]{Kupfer2015} and references therein as well as references in Table \ref{refl}\\
$^a$ paper II\\

%\qquad $^b$ \citet{baran19} \qquad $^c$ \citet{nemeth:2012,kawka:2015}

\newpage

\begin{table*}
\caption{Atmospheric parameters, luminosities and radii of the reflection effect candidates without known atmospheric parameters determined by SED fitting and \textit{Gaia} parallax.}
\label{sed}
\begin{tabular}{lllll}
\hline\hline
target & $T_{\rm eff,sed}$ & $\log{g}_{\rm sed,canonical}$ & $L_{\rm sed}$ & $R_{\rm sed}$ \\
& [K] & [cgs] & [L$_\odot$] & [R$_\odot$] \\\hline
2M0748+3042 & $33700^{+4400}_{-2900}$ & 5.78 & $21^{+14}_{-9}$ & $0.150^{+0.012}_{-0.015}$ \\
HE0505-3833 & $26500^{+1400}_{-1300}$ & 5.63 & $13.2^{+3.1}_{-2.6}$ & $0.175^{+0.008}_{-0.007}$ \\
MCT0049-3059 & $24300^{+700}_{-500}$ & 5.43 & $15.3^{+2.2}_{-1.8}$ & $0.221^{+0.010}_{-0.010}$ \\
TYC4542-482-1 & $16400^{+7900}_{-2400}$ & 5.15 & $8.7^{+2.6}_{-0.165}$ & $0.305^{+0.028}_{-0.052}$ \\
EC21390-2930 & $29000^{+12000}_{-5000}$ & 5.38 & $17^{+47}_{-14}$ & $0.241^{+0.026}_{-0.049}$ \\
2MASSJ18424506+6956202 & $37000^{+12000}_{-9000}$ & 5.7 & $17^{+67}_{-17}$ & $0.169^{+0.032}_{-0.030}$ \\
\hline
\end{tabular}
\end{table*}

\begin{table}[]
    \caption{Classification of \textit{TESS} targets with no variations}
    \label{no_var}
    \centering
    \begin{tabular}{lllccl}
    \hline\hline
    target & period & $K_1$ & $M_{2,\rm min}(M_{\rm sdB}=0.47\,\rm M_\odot$)& $M_{2,\rm min}(M_{\rm sdB}=0.4\,\rm M_\odot$) & signal-to-noise \\
    \hline
    \multicolumn{6}{c}{White dwarf companions (M dwarf excluded)}\\\hline
    KPD1930+2752 & 0.0950933 & 341 & 0.90 & 0.85 & ell \\
J083006+475150 & 0.1478000000 & 77.00 & 0.14 & 0.12 & 1\% \\
GALEXJ0805-1058 & 0.1737030000 & 29.20 & 0.05 & 0.04 & 0.04\% \\
J165404+303701 & 0.2535700000 & 126.10 & 0.32 & 0.29 & 0.4\% \\
HE0532-4503 & 0.2656000000 & 101.50 & 0.25 & 0.22 & 0.4\% \\
KUV16256+4034 & 0.4776000000 & 38.70 & 0.10 & 0.09 & 0.02\% \\
\hline
    \end{tabular}
\end{table}

\newpage
\begin{landscape}\scriptsize
\setlength{\tabcolsep}{1.5pt}

\begin{longtable}{l l l l l l l l l l l l l l}
\caption{All confirmed systems with light variations}
\label{tab}
\label{refl}\\
\hline\hline
TIC number & name & alternative & RA & DEC & classification & G & $B_P-R_P$& parallax & distance & G\_abs & reduced & period  & reference$^a$ \\
& & name & [deg] & [deg] & & [mag] & [mag] & [mas] & [kpc] & [mag] & pm & [h]  & \\\hline
 \multicolumn{14}{c}{reflection effect systems}\\\hline
4491131 & 2M0748+3042 & Ton287 & 117.233 & 30.713 & sdB & 14.046156 & -0.347797 & 1.334 & 0.750 & 4.672 & 10.473 & 5.53 & \citet{nemeth:2012}\\
14081239 & HE0505-3833 & HE0505-3833a & 76.745 & -38.488 & sdB & 14.145108 & -0.362259 & 1.228 & 0.815 & 4.591 & 8.454 & 7.73 & \citet{Donogue2013}\\
28762714 & FBS0145+363 & '' & 27.171 & 36.564 & sd & 15.462710 & -0.194446 & 3.786 & 0.264 & 8.354 & 12.404 & 3.03 & \citet{mickelian2008}\\
66398320 & MCT0049-3059 & PHL867 & 12.907 & -30.716 & sdB & 14.340959 & -0.392432 & 0.907 & 1.102 & 4.130 & 10.141 & 5.83 & \citet{Lamontagne2000}\\
\hline
\end{longtable}
\raggedright
$^a$ light variation analysis an/or atmospheric parameters or spectral classification

\end{landscape}
\end{appendix}

\end{document}